\documentclass[pre,twocolumn,showpacs,preprintnumbers,amsmath,amssymb]{revtex4}
\usepackage{graphicx,epsfig}
\usepackage{dcolumn}
\usepackage{bm}
\usepackage{psfrag}
\usepackage{plain}
\usepackage{tabularx}
\usepackage[latin1]{inputenc}
\usepackage{amsmath}
\begin{document}
\draft

\title{Dune formation on the present Mars}
\author{Eric J. R. Parteli$^1$ and Hans J. Herrmann$^{2,3}$}
\affiliation{1. Institut f\"ur Computerphysik, ICP, Universit\"at Stuttgart, Pfaffenwaldring 27, 70569 Stuttgart, Germany. \\ 2. Departamento de F\'{\i}sica, Universidade Federal do Cear\'a - 60455-760, Fortaleza, CE, Brazil. \\ 3. Computational Physics, IfB, ETH H\"onggerberg, HIF E 12, CH-8093, Z\"urich, Switzerland.}

\date{\today}

\begin{abstract}
We apply a model for sand dunes to calculate formation of dunes on Mars under the present Martian atmospheric conditions. We find that different dune shapes as those imaged by Mars Global Surveyor could have been formed by the action of sand-moving winds occuring on today's Mars. Our calculations show, however, that Martian dunes could be only formed due to the higher efficiency of Martian winds in carrying grains into saltation. The model equations are solved to study saltation transport under different atmospheric conditions valid for Mars. We obtain an estimate for the wind speed and migration velocity of barchan dunes at different places on Mars. From comparison with the shape of bimodal sand dunes, we find an estimate for the timescale of the changes in Martian wind regimes.

\end{abstract}


\maketitle


\section{\label{sec:introduction}Introduction}

Sand dunes are ubiquitous on Mars and provide evidence that sand-moving winds have once occured on the red planet. However, Martian dunes do not appear to be moving these days, and thus many authors suggested that dunes on Mars have been formed in the past, when the climate of Mars was much more earthlike \cite{Breed_et_al_1979}. On a planet where the present atmospheric density is almost hundred times lower than the Earth's, only winds one order of magnitude stronger than on Earth are able to transport sand and form dunes \cite{Greeley_et_al_1980,Iversen_and_White_1982}. Such strong winds occur in fact only occasionaly on Mars, during the strongest dust storms \cite{Moore_1985,Sullivan_et_al_2005}. Only once during the Viking mission, which operated for about six terrestrial years, could very little changes on Martian soils be detected after such a storm \cite{Arvidson_et_al_1983}. Moreover, calculations showed that the sand-moving winds that occurred at the Viking landing site lasted for not more than a few tens of seconds \cite{Arvidson_et_al_1983,Moore_1985}. Could Martian dunes have been formed by the action of such rarely occuring winds under the atmosphere of the present Mars? This is the question that motivates the present work.

The sand that forms dunes is transported by the wind through {\em{saltation}}, which consists of grains travelling in a sequence of ballistic trajectories and producing a {\em{splash}} of new ejected grains when impacting onto the ground \cite{Bagnold_1941}. Martian saltation has been studied in wind tunnel experiments and also in numerical simulations \cite{White_et_al_1976,White_1979}. Due to the thinner atmosphere of Mars and owing to the lower gravity $g=3.71$ m$/$s$^2$, which is nearly $1/3$ of the Earth's gravity, saltating grains on Mars travel longer and higher than their terrestrial counterparts \cite{White_1979}. Moreover, Martian grains also saltate with faster velocities than grains on our planet. As consequence, the grain-bed collisions or {\em{splash}} events on Mars are expected to be much larger than on Earth, due to the larger momentum transfered by the impacting grains to the sand bed \cite{Marshall_et_al_1998}. 

\begin{figure*}
\begin{center}
\vspace{0.5cm}
\includegraphics[width=1.0 \textwidth]{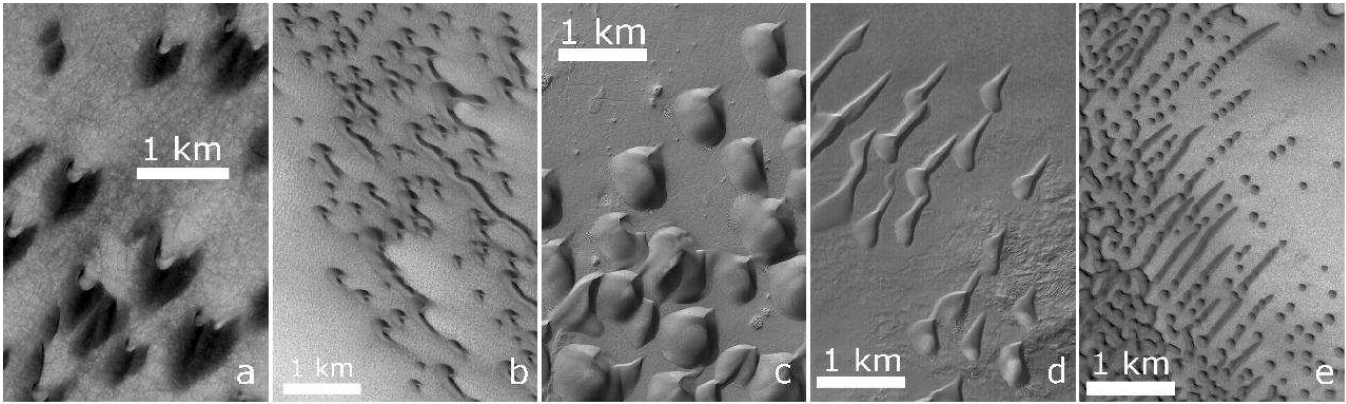}
\caption{Mars Global Surveyor (MGS), Mars Orbiter Camera (MOC) images of sand dunes on Mars (courtesy of NASA/JPL/MSSS). From the left to the right: {\bf{a.}} Barchan dunes at Arkhangelsky crater, near $41.0^{\circ}$S, $25.0^{\circ}$ W; {\bf{b.}} north polar dunes near $77.6^{\circ}$N, $103.6^{\circ}$ W; bimodal sand dunes near {\bf{c.}} $48.6^{\circ}$S, $25.5^{\circ}$ W; {\bf{d.}} $49.6^{\circ}$S, $352.9^{\circ}$ W, and {\bf{e.}} $76.4^{\circ}$N, $272.9^{\circ}$W.}
\label{fig:MOC_images}
\end{center}
\end{figure*}

What is not known is whether such highly energetic saltation events have been responsible for the formation of the enormous dunes observed in the images of Mars. In order to understand this, it is necessary to investigate sand transport at length scales comparable to the scale of dunes. Once saltation starts, the wind transfers momentum to accelerate the grains. Thus, the momentum of the air decreases as the flux of saltating particles increases (``feedback effect'' \cite{Owen_1964}). After a distance which is called ``saturation length'', the wind strength is just sufficient to sustain saltation, and the sand flux achieves saturation. In this manner, dunes that have length smaller than the saturation length will be continuously eroded due to increase of the sand flux and will disappear. In other words, the existence of a minimal dune size is related to the phenomenon of flux saturation, which could not be investigated from wind tunnel simulations of Martian saltation \cite{White_1979,Marshall_et_al_1998}. While the first wind tunnel simulating Martian conditions is a few meters long, the smallest dunes on Mars have length of the order of hundred meters (fig. \ref{fig:MOC_images}).

Recently, a successful modelling of the formation of sand dunes, which encompasses the main processes of saltation and accounts for flux saturation and the existence of a minimal dune size, has been achieved \cite{Sauermann_et_al_2001,Kroy_et_al_2002}. This model consists of a system of continuum equations in two space dimensions which reproduce the shape of terrestrial dunes, the wind profile and the sand flux and provide excellent quantitative agreement with measurements \cite{Sauermann_et_al_2003}. The dune model, which has been applied to study the interaction of dunes in a field \cite{Schwaemmle_and_Herrmann_2003} and the formation of parabolic dunes in the presence of vegetation \cite{Duran_and_Herrmann_2006a}, has become a powerful tool in the investigation of the large timescale processes involved in the formation of desert and coastal landscapes. 

In the present work, we apply the dune model to investigate whether dunes could be formed on the present Mars. Our aim to reproduce the shape of Martian dunes using the present Martian atmospheric conditions. The dune model has parameters of atmosphere, wind and sand, many of which are known for Mars and can be therefore used to solve the model equations. While most of the quantities controlling saltation can be calculated from the atmospheric density ${\rho}_{\mathrm{fluid}}$, gravity $g$, air viscosity $\eta$ and from the grain diameter $d$, there is one unknown quantity, which is related to the intensity of the Martian splash and must be determined from simulations. Moreover, the wind velocity that formed Martian dunes is also a parameter: it must be estimated from comparison with the shape of the dunes.

This paper is organized as follows. In the next Section we describe the dune model. The main equations are presented, as well as the relations used to calculate the microscopic quantities governing saltation on Mars. In Section \ref{sec:saltation} we calculate the average grain velocities, mean saltation height and sand flux on Mars from the known parameters, and the results are compared with wind tunnel predictions and with calculation results for saltation transport on Earth. Dunes on Mars are calculated in Section \ref{sec:dunes}. We begin with the simplest dune form, which is the barchan dune. First, we study the shape of the barchan dunes in the Arkhangelsky crater. We find an equation for the rate at which grains enter saltation, which can be used in the calculations of dunes under different atmospheric conditions. We then estimate the wind velocity on Mars and predict the migration velocity of Martian barchans. Next, we study the shape of Martian bimodal sand dunes and find an estimate for the timescale of changes in wind regimes on Mars. Conclusions are presented in Section \ref{sec:conclusions}.


\section{\label{sec:model}The dune model}

The dune model combines an analytical description of the average turbulent wind velocity field above the dune with a continuum saltation model which allows for saturation transients in the sand flux. In the model, avalanches on the slip face and flow separation at the dune lee are also taken into account. Here we give a brief presentation of the model and refer to Sauermann {\em{et al.}} \cite{Sauermann_et_al_2001} and Schw\"ammle and Herrmann \cite{Schwaemmle_and_Herrmann_2005} for the extensive derivation of the saltation transport equations. 


\subsection{Wind}

In the turbulent boundary layer, where sand transport takes place, the velocity of the wind $u(z)$ increases logarithmically with height $z$ above the flat ground:
\begin{equation}
u(z) = {\frac{u_{\ast}}{\kappa}}{\ln{\frac{z}{z_0}}} \label{eq:profile},
\end{equation}
where $\kappa = 0.4$ is the von K\'arm\'an constant, $u_{\ast}$ is the wind shear velocity, which is used to define the shear stress $\tau = {\rho}_{\mathrm{fluid}}{u_{\ast}^2}$, and $z_0$ is the aerodynamic roughness. $u_{\ast}$ and $z_0$ are two independent variables that can be determined from measurements of the wind velocity at different heights, as done for instance in the Mars Pathfinder Lander Wind Sock Experiment \cite{Sullivan_et_al_2000}. The aerodynamic roughness $z_0$ is larger than the surface roughness of the undisturbed sand bed, $z_0^{\mathrm{sand}}$, which is of the order of a few tens of microns and is due to the microscopic fluctuations of the sand bed when the grains are at rest \cite{Bagnold_1941}. A value of $z_0$ close to $1.0$ mm has been often reported from measurements of terrestrial saltation on a sand bed \cite{Pye_and_Tsoar_1990}, while on Mars $z_0$ is larger, around $1.0$ cm \cite{Sullivan_et_al_2000}.

A dune or a smooth hill introduces a perturbation in the shear stress whose Fourier-transformed components are calculated using the algorithm of Weng {\em{et al.}} \cite{Weng_et_al_1991}:
\begin{equation}
{\tilde{{{\hat{\tau}}}}}_x\!=\!{\frac{2\,h(k_x,k_y){k}_{x}^2}{|k|\,U^2(l)}}\!\left[\!{1\!+\!\frac{2{\ln({\cal{L}}|k_x|)\!+\!4{\epsilon}\!+\!1\!+\!{\mbox{i}}\,{\mbox{sign}}(k_x){\pi}}}{\ln{\left({l/z_0}\right)}}}\!\right] \label{eq:tau_x}
\end{equation}
and
\begin{equation}
{\tilde{{{\hat{\tau}}}}}_y={\frac{2\,h(k_x,k_y)k_xk_y}{|k|\,U^2(l)}}, \label{eq:tau_y}
\end{equation}
where the coordinate axes $x$ and $y$ are parallel, respectively, perpendicular to the wind direction, $k_x$ and $k_y$ are wave numbers, $|k|\!=\!\sqrt{k_{x}^2+k_{y}^2}$ and $\epsilon\!=\!0.577216$ (Euler's constant). ${\cal{L}}$ is the horizontal distance between the position of maximum height, $H_{\mathrm{max}}$, and the position of the windward side where the height is $H_{\mathrm{max}}/2$ \cite{Weng_et_al_1991}. $U(l)\!=\!u(l)/u(h_{\mathrm{m}})$ is the undisturbed wind velocity at height $l\!=\!{2{\kappa}^2{\cal{L}}}/{\ln{{l}/z_0}}$ normalized by the velocity at the reference height $h_{\mathrm{m}}\!=\!{\cal{L}}/{{\sqrt{{\log{{\cal{L}}/z_0}}}}}$, which separates the middle and upper flow layers \cite{Weng_et_al_1991}. The shear stress in the direction $i$ ($i\!=\!x,y$) is then given by ${\vec{{\tau}}}_i\!=\!{\hat{{i}}}\left[{{\tau}_0{(1+{{\hat{\tau}}}_i)}}\right]$, where ${\tau}_0$ is the undisturbed shear stress over the flat ground.

Sand transport occurs if $u_{\ast}$ exceeds a threshold velocity for entrainment, $u_{{\ast}{\mathrm{ft}}}$, which depends on $d$, $g$, ${\rho}_{\mathrm{fluid}}$, on the grain density ${\rho}_{\mathrm{grain}}$ and also on the packing of the grains \cite{Shields_1936}. Indeed, the wind velocity may even decrease to values lower than $u_{{\ast}{\mathrm{ft}}}$, and still saltation can be sustained, once initiated. This is because the splash is the most important mechanism of sand entrainment during saltation \cite{Bagnold_1941}. The wind strength, however, cannot be lower than the impact threshold velocity $u_{{\ast}{\mathrm{t}}}$, which defines the threshold shear stress ${\tau}_{\mathrm{t}} = {\rho}_{\mathrm{fluid}}u_{{\ast}{\mathrm{t}}}^2$ and is around $80\%$ of $u_{{\ast}{\mathrm{ft}}}$. Saltation ceases if $u_{\ast} < u_{{\ast}{\mathrm{t}}}$, and therefore the impact threshold velocity is the essential threshold velocity for aeolian sand transport.


\subsection{Continuum saltation model}

The wind shear velocity computed above is used to calculate the sand flux with the model derived in Sauermann {\em{et al.}} \cite{Sauermann_et_al_2001}. The fundamental idea of the model is to consider the bed-load as a thin fluid-like granular layer on top of an immobile sand bed. 

The sand bed represents an open system which can exchange grains with the moving saltation layer, for which the erosion rate $\Gamma(x,y)$ at any position ($x,y$) represents a source term. Mass conservation yields that the local change in the flux balances the erosion rate: 
\begin{equation}
{\vec{\nabla}}{\cdot}({\rho}{\vec{v}}) = {\Gamma}(x,y), \label{eq:balance}
\end{equation}
where ${\rho}(x,y)$ is the density of grains in the saltation layer, and ${\vec{v}}(x,y)$ is the average local velocity of the saltating grains, whereas stationary condition has been assumed (${\partial}/{\partial}t = 0$) since the time scale of the surface evolution is several orders of magnitude larger than the typical values of transient time of the saltation flux. The erosion rate is the difference between the vertical flux of ejected grains and the vertical flux ${\phi}$ of grains impacting onto the bed: 
\begin{equation}
{\Gamma} = {\phi}(n-1), \label{eq:erosion_rate}
\end{equation}
where $n$ is the average number of splashed grains. The vertical flux $\phi$ is defined as $\phi = {\rho}|{\vec{v}}|/{\ell}$, where $\ell$ is the average saltation length. Due to the multiplicative process inherent in the splash events, the number of saltating grains first increases exponentially. However, the ``feedback effect'' leads to a decrease of the momentum of the air as the population of grains increases. At saturation, the air shear stress ${\tau}_{\mathrm{a}}$ decreases to the threshold ${\tau}_{\mathrm{t}}$, while the flux of grains increases to its maximum value. The number of ejecta compensates the number of impacting grains: $n = 1$ at saturation. In this manner, we write $n$ as a function $n({\tau}_{\mathrm{a}}/{\tau}_{\mathrm{t}})$ with $n(1) = 1$. Expansion of $n$ into a Taylor series up to the first order term at the threshold yields
\begin{equation}
n = 1 + {\tilde{\gamma}}{\left({{\frac{{\tau}_{\mathrm{a}}}{{\tau}_{\mathrm{t}}}}-1}\right)} \label{eq:splash},
\end{equation} 
where ${\tilde{\gamma}} = {\mbox{d}}n/{{\mbox{d}}({\tau}_{\mathrm{a}}/{\tau}_{\mathrm{t}})}$, the {\em{entrainment rate}} of grains into saltation, determines how fast the system reaches saturation \cite{Sauermann_et_al_2001}. Inserting eq. (\ref{eq:splash}) into eq. (\ref{eq:erosion_rate}), and substituting this into eq. (\ref{eq:balance}), we obtain a differential equation for the sand flux.

However, the air shear stress ${\tau}_{\mathrm{a}}$ within the saltation layer is written as ${\tau}_{\mathrm{a}} = \tau - {\tau}_{\mathrm{g}}$, where ${\tau}_{\mathrm{g}}$ is the contribution of the grains to the total shear stress at the ground \cite{Sauermann_et_al_2001}. The ``grain born'' shear stress is calculated as ${\tau}_{\mathrm{g}} = {\phi}{\Delta}v_{\mathrm{hor}}={\rho}|{\vec{v}}|{\Delta}v_{\mathrm{hor}}/{\ell}$, where ${\Delta}v_{\mathrm{hor}} = v_{\mathrm{hor}}^{\mathrm{imp}}-v_{\mathrm{hor}}^{\mathrm{eje}}$ gives the difference between the horizontal velocities (in the direction of the flow) of the grains at the moment of impact, $v_{\mathrm{hor}}^{\mathrm{imp}}$, and at the moment of ejection, $v_{\mathrm{hor}}^{\mathrm{eje}}$. The equation of mass conservation can be written, then, in the following manner:
\begin{equation}
\vec{\nabla} \cdot ({\rho}\vec{v}) = {\frac{{\rho}|{\vec{v}}|}{\ell}}{{\tilde{\gamma}}}{\frac{{\tau} - {\tau}_{\mathrm{t}}}{{\tau}_{\mathrm{t}}}} {\left({1 - {\rho}|{\vec{v}}|{\frac{{\Delta}v_{\mathrm{hor}}/{\ell}}{{\tau} - {\tau}_{\mathrm{t}}}}}\right)}. \label{eq:balance_2}
\end{equation}
On the other hand, the mean saltation length $\ell$ is defined in terms of the grain velocity $|{\vec{v}}|$ and of the initial vertical velocity, $v_z^{\mathrm{eje}}$ \cite{Sauermann_et_al_2001}: $\ell = v_z^{\mathrm{eje}}(2|{\vec{v}}|/g)$, where $v_z^{\mathrm{eje}}$ is related to the gain in horizontal velocity of the grains, ${\Delta}v_{\mathrm{hor}}$, through an effective restitution coefficient for the grain-bed interaction, $\alpha$ \cite{Sauermann_et_al_2001}, which is defined as
\begin{equation}
\alpha = \frac{v_z^{\mathrm{eje}}}{{{\Delta}v_{\mathrm{hor}}}} = \frac{v_z^{\mathrm{eje}}}{v_{\mathrm{hor}}^{\mathrm{imp}} - v_{\mathrm{hor}}^{\mathrm{eje}}}. \label{eq:alpha_definition}
\end{equation}
In this manner, the mean saltation length is written as 
\begin{equation}
{\ell} = v_z^{\mathrm{eje}}{\frac{2|{\vec{v}}|}{g}} = {\frac{\alpha}{{\Delta}v_{\mathrm{hor}}}}{\frac{2|{\vec{v}}|}{g}} = {\frac{1}{r}}{\left[{\frac{2{|{\vec{v}}|}^2{\alpha}}{g}}\right]}, \label{eq:saltation_length}
\end{equation}
where 
\begin{equation}
r = \frac{|{\vec{v}}|}{{{\Delta}v_{\mathrm{hor}}}} \label{eq:r_definition}
\end{equation}
is the constant of proportionality between the average grain velocity ${|{\vec{v}}|}$ and the difference between impact and ejection velocity of the grains, ${\Delta}v_{\mathrm{hor}}$.

The velocity of the saltating grains, ${\vec{v}}$, is determined from the balance between three forces: (i) the drag force acting on the grains; (ii) the bed friction which yields the loss of momentum when the grains impact onto the ground, and (iii) the downhill force, which acts on the saltation layer in the presence of bed slopes. To calculate the grain velocity, we need to take into account the modification of the air flow due to the presence of the saltating grains. However, the model equations do not account for the complex velocity distribution within the saltation layer. Instead, a reference height $z_1$ is taken, between the ground at the roughness height $z_0^{\mathrm{sand}}$ and the mean saltation height $z_{\mathrm{m}}$, at which the ``effective'' wind velocity, ${\vec{u}}_{\mathrm{eff}}$, is calculated \cite{Sauermann_et_al_2001}.

Finally, a useful approximation is employed, which simplifies the equations in significant manner and leads to only a negligible error \cite{Sauermann_et_al_2001}. In the model, the velocity ${\vec{u}}_{\mathrm{eff}}$ is calculated in the {\em{steady state}}, i.e. it is the {\em{reduced}} wind velocity within the saltation layer at saturation. In geomorphological applications, the sand flux is nearly everywhere saturated, with exception of those places where external variables change discontinuously, as for instance at a flow-separation, which occurs at the dune brink, or at a phase boundary bedrock$/$sand which occurs at the windward foot of a barchan dune. Therefore, we can replace the density ${\rho}$ which appears in the expression for the grain born shear stress ${\tau}_{\mathrm{g}}$ with the saturated density ${\rho}_{\mathrm{s}} = ({\tau}-{\tau}_{\mathrm{t}}){\ell}/{\Delta}v_{\mathrm{hor}}$ \cite{Sauermann_et_al_2001}. The following expression is obtained, in this manner, for ${\vec{u}}_{\mathrm{eff}}$:
\begin{equation}
{\vec{u}}_{\mathrm{eff}} = {\frac{u_{{\ast}{\mathrm{t}}}}{\kappa}}{\left\{{{\ln{\frac{z_1}{z_0^{\mathrm{sand}}}}} + 2{\left[{{\sqrt{1 + {\frac{z_1}{z_{\mathrm{m}}}}{\left({{\frac{u_{\ast}^2}{u_{{\ast}{\mathrm{t}}}^2}}-1}\right)}}-1}}\right]}}\right\}}{\frac{{\vec{u}}_{\ast}}{|{\vec{u}}_{\ast}|}}. \label{eq:u_eff}
\end{equation}
The grain velocity, $\vec{v}$, is, next, calculated numerically from the equation \cite{Sauermann_et_al_2001}:
\begin{equation}
\frac{3}{4}{\frac{{\rho}_{\mathrm{fluid}}}{{\rho}_{\mathrm{grain}}}}{\frac{C_{\mathrm{d}}}{d}}({\vec{u}}_{\mathrm{eff}} - \vec{v})|{\vec{u}}_{\mathrm{eff}}-\vec{v}| - {\frac{g{\vec{v}}}{2{\alpha}{|\vec{v}|}}} - g{\vec{\nabla}}h = 0, \label{eq:velocity}
\end{equation}
where $C_{\mathrm{d}}$ is the drag coefficient, and ${\vec{u}}_{\mathrm{eff}}$ is calculated with eq. (\ref{eq:u_eff}). In this manner, the grain velocity obtained from eq. (\ref{eq:velocity}) is in fact the {\em{average grain velocity}} at the steady state, ${\vec{v}}_{\mathrm{s}}$, since ${\vec{u}}_{\mathrm{eff}}$ is the reduced wind velocity after flux saturation has been achieved \cite{Sauermann_et_al_2001}. Moreover, the mean saltation length (eq. (\ref{eq:saltation_length})) is also computed using $|{\vec{v}}| = |{\vec{v}}_{\mathrm{s}}|$ in the stationary condition where the sand flux is in equilibrium. The average grain velocity ${\vec{v}}_{\mathrm{s}}$ can be now substituted into eq. ({\ref{eq:balance_2}}), which is then written in terms of the {\em{sand flux}} 
\begin{equation}
{\vec{q}} = {{\rho}{\vec{v}}_{\mathrm{s}}}, \label{eq:sand_flux}
\end{equation}
and the {\em{saturated sand flux}} $q_{\mathrm{s}} = {\rho}_{\mathrm{s}}|{{\vec{v}}_{\mathrm{s}}}|$, or
\begin{equation}
q_{\mathrm{s}} = {\frac{2{\alpha}|{{\vec{v}}_{\mathrm{s}}}|}{g}}{({{\tau} - {\tau}_{\mathrm{t}}})} = {\frac{2{\alpha}|{{\vec{v}}_{\mathrm{s}}}|}{g}}{u_{{\ast}{\mathrm{t}}}^2}{\left[{{\left({{u_{\ast}}/u_{{\ast}{\mathrm{t}}}}\right)}^2 - 1}\right]}. \label{eq:saturated_flux}
\end{equation}
The resulting equation for the sand flux is a differential equation that contains the saturated flux $q_{\mathrm{s}}$ at the steady state,
\begin{equation}
{\vec{\nabla}}{\cdot}{\vec{q}}  = {\frac{1}{{\ell}_{\mathrm{s}}}}|{{\vec{q}}}|{\left({1 - {\frac{|{\vec{q}}|}{q_{\mathrm{s}}}}}\right)}, \label{eq:differential_q}
\end{equation}
where ${\ell}_{\mathrm{s}} = [{\ell}/{\tilde{\gamma}}]{\tau}_{\mathrm{t}}/({\tau}-{\tau}_{\mathrm{t}})$ is the {\em{saturation length}}, which contains the information of the saturation transient of the sand flux. Using eq. (\ref{eq:saltation_length}), ${\ell}_{\mathrm{s}}$ may be written as
\begin{equation}
{\ell}_{\mathrm{s}} = {\frac{1}{\tilde{\gamma}}}{\left[{\frac{\ell}{{\left({{u_{\ast}}/u_{{\ast}{\mathrm{t}}}}\right)}^2 - 1}}\right]} = {\frac{1}{{\gamma}}}{\left[{\frac{2{|{\vec{v}}_{\mathrm{s}}|}^2{\alpha}/g}{{\left({{u_{\ast}}/u_{{\ast}{\mathrm{t}}}}\right)}^2 - 1}}\right]}, \label{eq:saturation_length}
\end{equation}
where we defined
\begin{equation}
\gamma = r{\tilde{\gamma}} = {\frac{|{\vec{v}}_{\mathrm{s}}|}{{{\Delta}v_{\mathrm{hor}}}}}\left[{\frac{{\mbox{d}}n}{{{\mbox{d}}({\tau}_{\mathrm{a}}/{\tau}_{\mathrm{t}})}}}\right]. \label{eq:r_gamma} 
\end{equation}


\subsection{Surface evolution}

The change in the surface is computed using the flux calculated with eq. (\ref{eq:differential_q}). The surface is eroded wherever the sand flux increases in the direction of wind flow, and sand deposition takes place if the flux decreases. The time evolution of the topography $h(x,y,t)$ is given by the mass conservation equation:
\begin{equation}
\frac{{\partial}h}{{\partial}t} = - {\frac{1}{{\rho}_{\mathrm{sand}}}}{\vec{\nabla}}{\cdot}{\vec{q}}, \label{eq:time_evolution}
\end{equation}
where ${\rho}_{\mathrm{sand}} = 0.62 {\rho}_{\mathrm{grain}}$ is the mean density of the immobile sand which constitutes the sand bed \cite{Sauermann_et_al_2001}. If sand deposition leads to slopes that locally exceed the angle of repose ${\theta}_{\mathrm{r}} \approx 34^{\circ}$, the unstable surface relaxes through avalanches in the direction of the steepest descent. Avalanches are assumed to be instantaneous since their time scale is negligible in comparison with the time scale of the dune motion. At the brink of the dune, which represents a sharp edge, there occurs flow separation. In the model, the separation streamlines are introduced at the dune lee as described in details in Kroy {\em{et al.}} \cite{Kroy_et_al_2002}. Each streamline is fitted by a third order polynomial connecting the brink with the ground at the reattachment point \cite{Kroy_et_al_2002}, and defining the ``separation bubble'', in which the wind and the flux are set to zero. 

The dune model can be sketched as follows: 
\begin{enumerate}
\item the shear stress over the surface is calculated with the algorithm of Weng {\em{et al.}} \cite{Weng_et_al_1991}, using eqs. (\ref{eq:tau_x}) and (\ref{eq:tau_y}); 
\item from the shear stress, the sand flux is calculated using eq. (\ref{eq:differential_q}), where the saturation length ${\ell}_{\mathrm{s}}$ and the saturated sand flux $q_{\mathrm{s}}$ are calculated from expressions (\ref{eq:saturation_length}) and ({\ref{eq:saturated_flux}}), respectively; 
\item the change in the surface height is computed from mass conservation (eq. (\ref{eq:time_evolution})) using the calculated sand flux; and
\item  avalanches occur wherever the inclination exceeds $34^{\circ}$, then the slip face is formed and the separation streamlines are introduced as described in Kroy {\em{et al.}} \cite{Kroy_et_al_2002}.
\end{enumerate}
Calculations are performed using open boundaries with a constant influx of sand, $q_{\mathrm{in}}$, at the inlet. The influx is interpreted as the average interdune flux in a dune field, which is typically between $10$ and $40\%$ of the maximum flux $q_{\mathrm{s}}$ \cite{Fryberger_et_al_1984}, and is considered, for simplicity, homogeneous along the $y$ axis (perpendicular to sand transport). The model is evaluated by performing steps 1) through 4) computationally in a cyclic manner. 


\subsection{Model parameters}

The following quantities are needed in order to solve the model equations: the atmospheric density ${\rho}_{\mathrm{fluid}}$, gravity $g$, grain diameter $d$ and density ${\rho}_{\mathrm{grain}}$, whose values are found in the literature and are discussed in the next Section; the impact threshold velocity for saltation, $u_{{\ast}{\mathrm{t}}}$, and the drag coefficient $C_{\mathrm{d}}$; the effective restitution coefficient $\alpha$ and the heights $z_{\mathrm{m}}$, $z_1$ and $z_0^{\mathrm{sand}}$; $\gamma$ (eq. (\ref{eq:r_gamma})) and the wind shear velocity $u_{\ast}$. 

The {\em{impact threshold velocity}} $u_{{\ast}{\mathrm{t}}}$ is about $80\%$ of the threshold for aeolian entrainment, $u_{{\ast}{\mathrm{ft}}}$ \cite{Bagnold_1941}, which in turn is calculated as in Iversen and White \cite{Iversen_and_White_1982}. This leads to the following equation for $u_{{\ast}{\mathrm{t}}}$:
\begin{equation}
u_{{\ast}{\mathrm{t}}}= 0.8\,A\,\sqrt{{\frac{{({\rho}_{\mathrm{grain}}-{\rho}_{\mathrm{fluid}})}gd}{{\rho}_{\mathrm{fluid}}}}}, \label{eq:u_t}
\end{equation}
where $A$ is called the Shields parameter, which depends on the shape and sorting of the grains and on the angle of internal friction \cite{Shields_1936}. The Shields parameter is calculated as in Iversen and White \cite{Iversen_and_White_1982}:
\begin{equation}
A = 0.129 {\left[{{\frac{{\left({1 + 6.0 \times 10^{-7}/{{{\rho}_{\mathrm{grain}}}gd^{2.5}}}\right)}^{0.5}}{{\left({1.928{\mbox{Re}}_{{\ast}{\mathrm{ft}}}^{0.092}}-1\right)}^{0.5}}}}\right]} \label{eq:Shields_parameter_a}
\end{equation} 
for $0.03 \leq {\mbox{Re}}_{{\ast}{\mathrm{ft}}} \leq 10$ and 
\begin{eqnarray}
A  && = 0.129 {\left({1 + 6.0 \times 10^{-7}/{{{\rho}_{\mathrm{grain}}}gd^{2.5}}}\right)}^{0.5} \nonumber \\ & & \cdot {\left\{{1 - 0.0858 \exp{\left[{-0.0617({\mbox{Re}}_{{\ast}{\mathrm{ft}}}-10)}\right]}}\right\}}  \label{eq:Shields_parameter_b}
\end{eqnarray}
for ${\mbox{Re}}_{{\ast}{\mathrm{ft}}} \geq 10$, where ${\mbox{Re}}_{{\ast}{\mathrm{ft}}}$ is the friction Reynolds number ${\mbox{Re}}_{{\ast}{\mathrm{ft}}} \equiv u_{{\ast}{\mathrm{ft}}}d/{\nu}$, and the constant $6.0 \times 10^{-7}$ has units of ${\mbox{kg}}{\cdot}{\mbox{m}}^{0.5}{\cdot}{\mbox{s}}^{-2}$, while all other numbers are dimensionless. The kinematic viscosity $\nu$ is defined as ${\eta}/{\rho}_{\mathrm{fluid}}$, where $\eta$ is the dynamic viscosity. We notice that in contrast to $\nu$, $\eta$ depends only on the atmospheric temperature and composition. 

The {\em{drag coefficient}} $C_{\mathrm{d}}$ is a function of the Reynolds number Re. Jim\'enez and Madsen \cite{Jimenez_and_Madsen_2003} calculated the drag coefficient $C_{\mathrm{d}}$ of a particle falling with settling velocity $v_{\mathrm{f}}$ from the balance between the gravitational force and the drag resistance of the fluid. To adapt the formula of Jim\'enez and Madsen \cite{Jimenez_and_Madsen_2003} --- which is valid for Re within the range $0.2 < {\mbox{Re}} < 127$ --- to grain saltation, we consider the balance between the fluid drag on the grains in the saltation layer and the bed friction that compensates the grain-born shear stress at the surface. This leads to the following equation \cite{Jimenez_and_Madsen_2003,Duran_and_Herrmann_2006b}:
\begin{equation}
C_{\mathrm{d}} = {\frac{4}{3}}{\left({A_{\mathrm{d}} + \frac{B_{\mathrm{d}}}{S}}\right)}^2, \label{eq:C_d}
\end{equation}
where 
\begin{equation}
S = \frac{d}{4{\nu}}{\sqrt{{{\frac{1}{2{\alpha}}}}{\left[{\frac{{({\rho}_{\mathrm{grain}}-{\rho}_{\mathrm{fluid}})}gd}{{\rho}_{\mathrm{fluid}}}}\right]}}} \label{eq:S}
\end{equation}
is called the fluid-sediment parameter and $A_{\mathrm{d}}$ and $B_{\mathrm{d}}$ are constants that contain information about the sediment shape factor and roundness. In this manner, the drag coefficient for saltating particles is the same as in Jim\'enez and Madsen \cite{Jimenez_and_Madsen_2003} but with the quantity $S$ (eq. ({\ref{eq:S}})) corrected by a factor $1/{\sqrt{2{\alpha}}}$. Furthermore, Jim\'enez and Madsen \cite{Jimenez_and_Madsen_2003} suggested to use $A_{\mathrm{d}}=0.95$ and $B_{\mathrm{d}}=5.12$ for typical applications when particle's shape and roundness are not known. 

The parameters $\alpha$, $z_{\mathrm{m}}$ and $z_1$ are computed using the equations obtained by Dur\'an and Herrmann \cite{Duran_and_Herrmann_2006b}. These equations have been obtained from comparison with wind tunnel data of Rasmussen {\em{et al.}} \cite{Rasmussen_et_al_1996} and Iversen and Rasmussen \cite{Iversen_and_Rasmussen_1999}, and allow to calculate the model parameters for saltation in different physical environments. Here we just display the equations and refer to the original reference \cite{Duran_and_Herrmann_2006b} for detailed explanation.

The equations for the model parameters are scaling relations that incorporate the timescale 
\begin{equation}
t_{\nu} \equiv {({\nu}/g^2)}^{1/3} \label{eq:t_nu}
\end{equation}
and the lengthscale
\begin{equation}
{\ell}_{\nu} \equiv {\left[{{\frac{{\nu}^2{\rho}_{\mathrm{fluid}}}{A^2g({\rho}_{\mathrm{grain}}-{\rho}_{\mathrm{fluid}})}}}\right]}^{1/3}. \label{l_nu}
\end{equation}
The reference height $z_1$, at which the effective wind velocity $u_{\mathrm{eff}}$ is calculated, is given by the equation 
\begin{equation}
z_1 = 35{\ell}_{\nu}. \label{eq:z_1}
\end{equation}
The height $z_1$ is between the mean saltation height 
\begin{equation}
z_{\mathrm{m}} = 14u_{{\ast}{\mathrm{t}}}t_{\nu}, \label{eq:z_m}
\end{equation}
and the surface roughness 
\begin{equation}
z_0^{\mathrm{sand}} = d/20. \label{eq:z0_sand}
\end{equation}
The last equation gives intermediate values between $d/30$ \cite{Bagnold_1941} and $d/8$ \cite{Andreotti_2004}. Finally, the effective restitution coefficient $\alpha$ (eq. (\ref{eq:alpha_definition})) is simply calculated with the formula \cite{Duran_and_Herrmann_2006b}
\begin{equation}
\alpha = 0.17d/{\ell}_{\nu}. \label{eq:alpha}
\end{equation}
{\em{Summary}} --- With the equations presented in this Section, the model parameters can be calculated from the following quantities: ${\rho}_{\mathrm{fluid}}$, $g$, $d$, ${\rho}_{\mathrm{grain}}$, and from the viscosity ${\eta}$. Thus, these quantities are, together with $\gamma$ (eq. (\ref{eq:r_gamma})) and with the shear velocity $u_{\ast}$, the only parameters of the model. The threshold velocity $u_{{\ast}{\mathrm{t}}}$ is obtained with eq. ({\ref{eq:u_t}}), the drag coefficient $C_{\mathrm{d}}$ is given by eq. (\ref{eq:C_d}), while eqs. (\ref{eq:z_1}) --- (\ref{eq:alpha}) are used to obtain $z_1$, $z_{\mathrm{m}}$, $z_0^{\mathrm{sand}}$ and $\alpha$.


\section{\label{sec:saltation}Saltation transport on Mars}

Since the quantities governing saltation are functions of the atmospheric conditions, we expect saltation to be different depending on the location on Mars. The reason is that the average atmospheric pressure and temperature may vary within an extremely wide range compared to the terrestrial case. In this Section, we estimate the average trajectories of saltating grains and the sand flux under different atmospheric conditions on Mars. The results presented in this Section are then used in the next Sections, to calculate formation of dunes on Mars.


\subsection{Martian atmosphere}

The Mars Global Surveyor Radio Science (MGSRS) Team has provided valuable atmospheric data of Mars \cite{MGSRS}. In particular, the temperature $T$ and the pressure $P$ near the surface have been systematically measured in many locations of Mars. We use the ideal gas equation to calculate the local atmospheric density, ${\rho}_{\mathrm{fluid}}$, from the MGSRS pressure and temperature data. An atmosphere of $100\%$ CO$_2$ is considered. Furthermore, the dynamic viscosity $\eta$ of the Martian atmosphere is a function of the temperature $T$, and is calculated using the Sutherland's formula \cite{Crane_1988}:
\begin{equation}
\eta = {\eta}_0{\left[{\frac{T_0 + C}{T + C}}\right]}{(T/T_0)}^{3/2}, \label{eq:eta}
\end{equation}
where for CO$_2$ we have ${\eta}_{\mathrm{0}} = 1.48{\cdot}10^{-5}$ kg$/$m${\cdot}$s, $C = 240$, $T_0 = 293.15$ K \cite{Crane_1988}. Finally, the kinematic viscosity ${\nu}$ is calculated with the equation $\nu = \eta/{{\rho}_{\mathrm{fluid}}}$.


\subsection{Particle size of Martian dunes}

Edgett and Christensen \cite{Edgett_and_Christensen_1991} have used thermal inertia data to obtain the grain diameter of dunes in intra-crater fields of dark dunes on Mars. They found that the average grain diameter of Martian dunes is $d=500 \pm 100$ ${\mu}$m, which is coarser than the mean diameter of terrestrial dune grains, $250$ ${\mu}$m \cite{Pye_and_Tsoar_1990}. The value of mean grain diameter $d=500$ ${\mu}$m as measured by Edgett and Christensen \cite{Edgett_and_Christensen_1991} for dunes on Mars is used in the calculations of the present work, while $d=250$ ${\mu}$m is considered for terrestrial dunes. Furthermore, we take the density ${\rho}_{\mathrm{grain}} = 3200$ and $2650$ kg$/$m$^3$ for Martian \cite{Fenton_et_al_2003} and terrestrial \cite{Bagnold_1941} grains, respectively.

Why is the sand of Martian dunes coarser than the sand of Earth's dunes? There is a critical value of the diameter $d$, below which the particle remains suspended in the atmosphere. The critical diameter depends on the vertical fluctuating component of the wind velocity $u^{\prime}$ \cite{Tsoar_and_Pye_1987}. If the standard deviation of $u^{\prime}$, which scales with the wind friction speed, is larger than the settling velocity of the grain, $v_{\mathrm{f}}$, then the particle will remain suspended. The falling velocity $v_{\mathrm{f}}$ is obtained from the equilibrium between the gravitational force and the fluid drag \cite{Jimenez_and_Madsen_2003}. Particles for which the ratio $v_{\mathrm{f}}/u_{{\ast}{\mathrm{ft}}}$ is smaller than $1.0$ enter suspension \cite{Tsoar_and_Pye_1987}. In this manner, a critical grain diameter of $210$ ${\mu}$m is obtained for Mars, while on Earth the critical value is about $52$ ${\mu}$m \cite{Edgett_and_Christensen_1991}. 

The critical diameter obtained in this manner for Mars appears inconsistent with the observation of Martian aeolian ripples composed of grains of diameter around $100$ ${\mu}$m \cite{Sullivan_et_al_2005,Claudin_and_Andreotti_2006}. In order to calculate the transition suspension/saltation, we use the ``falling'' velocity $v_{\mathrm{f}}^{\ast}$ obtained from the equilibrium between the fluid drag and the bed friction ${\tau}_{\mathrm{g}}$. This ``falling'' velocity is given by the equation
\begin{equation}
v_{\mathrm{f}}^{\ast} = \sqrt{{\frac{4}{3C_{\mathrm{d}}}}{\left[{\frac{{({\rho}_{\mathrm{grain}}-{\rho}_{\mathrm{fluid}})}gd}{{\rho}_{\mathrm{fluid}}}}\right]}}, \label{eq:v_f}
\end{equation}
in which the value of $C_{\mathrm{d}}$ (eq. (\ref{eq:C_d})) differs from the drag coefficient of a vertically falling grain by a factor of the order of $1/2{\alpha}$, where $\alpha$ is given by eq. (\ref{eq:alpha}). 

\begin{figure}[htpb]
\begin{center}
\vspace{0.5cm}
\includegraphics[width=0.95 \columnwidth]{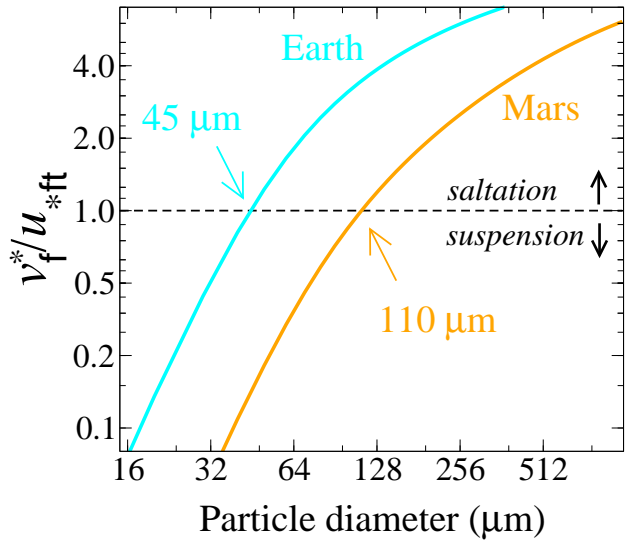}
\caption{Ratio between the falling velocity $v_{\mathrm{f}}^{\ast}$ and the threshold friction speed for saltation, $u_{{\ast}{\mathrm{ft}}}$, calculated for Mars and for Earth as function of the grain diameter $d$. Dashed line indicates the saltation$/$suspension transition at $v_{{\mathrm{f}}}^{\ast}/u_{{\ast}{\mathrm{ft}}}=1.0$, and intercepts the Martian (terrestrial) continuous line at $d=110{\mu}$m ($d=45{\mu}$m).}
\label{fig:suspension_threshold}
\end{center}
\end{figure}

Figure \ref{fig:suspension_threshold} shows the ratio $v_{\mathrm{f}}^{\ast}/u_{{\ast}{\mathrm{ft}}}$ as function of the grain diameter $d$, calculated using parameters for Earth and for Mars, where we used the nominal pressure $P = 6.0$ mbar and temperature $T = 200$ K for Mars. In this figure, the threshold shear velocity for saltation, $u_{{\ast}{\mathrm{ft}}} = 1.25\,u_{{\ast}{\mathrm{t}}}$, is calculated using eq. (\ref{eq:u_t}), while $v_{\mathrm{f}}^{\ast}$ is calculated with eq. (\ref{eq:v_f}). The dashed line indicates the transition value $v_{\mathrm{f}}^{\ast}/u_{{\ast}{\mathrm{ft}}}=1.0$.

As we can see from fig. \ref{fig:suspension_threshold}, particles with diameter smaller than $45$ ${\mu}$m enter suspension on Earth, while the critical value of $d$ on Mars is around $110$ ${\mu}$m. This value is larger than the terrestrial one, but is smaller than the one obtained in previous calculations \cite{Edgett_and_Christensen_1991}. Furthermore, we see that the ratio between the measured average grain size of dunes ($500$ and $250$ ${\mu}$m on Mars and on Earth, respectively) and the critical diameter obtained in fig. \ref{fig:suspension_threshold} is around $5.0$ on both planets. In fact, pure saltation is expected to occur only if the falling velocity is larger than $2.5\,u_{{\ast}{\mathrm{t}}}$, which explains why the sand of dunes is effectively much larger than the critical diameter \cite{Edgett_and_Christensen_1991}. 


\subsection{Saltation trajectories and sand flux}

The model parameters that govern the grain trajectories are the average saltation height, $z_{\mathrm{m}}$ (eq. (\ref{eq:z_m})); the reference height $z_1$ (eq. (\ref{eq:z_1})) at which the effective wind velocity $u_{\mathrm{eff}}$ is calculated; the drag coefficient, $C_{\mathrm{d}}$ (eq. (\ref{eq:C_d})); and the effective restitution coefficient, $\alpha$ (eq. (\ref{eq:alpha})). From $d=500$ ${\mu}$m, we obtain the surface roughness $z_0^{\mathrm{sand}} = 25$ ${\mu}$m (eq. (\ref{eq:z0_sand})). The saturated sand flux $q_{\mathrm{s}}$ (eq. (\ref{eq:saturated_flux})) is function of the wind shear velocity $u_{\ast}$ and further depends on the saturation velocity of the saltating grains, $v_{\mathrm{s}}=|{\vec{v}}_{\mathrm{s}}|$, which is calculated in eq. (\ref{eq:velocity}). The quantities controlling saltation on Mars are calculated in table \ref{tab:parameters_Mars}. 

In table \ref{tab:parameters_Mars}, the model parameters have been calculated using different values of pressure $P$ and temperature $T$ valid for Mars. We see that the minimal friction speed for saltation, $u_{{\ast}{\mathrm{t}}}$, on Mars may vary by a factor of 2. We note that ranges of $P$ and $T$ even wider than the ones studied here may occur on Mars \cite{MGSRS}. Moreover, we calculate $v_{\mathrm{s}}$ and $q_{\mathrm{s}}$ in table \ref{tab:parameters_Mars} using a constant value of $u_{\ast}/u_{{\ast}{\mathrm{t}}} = 1.5$, since this is a representative value for saltation on Earth \cite{Fryberger_and_Dean_1979}. The corresponding values calculated for Earth are shown in the last row of table \ref{tab:parameters_Mars}, where we used $d=250$ ${\mu}$m, density ${\rho}_{\mathrm{fluid}} = 1.225$ kg$/$m$^3$ and viscosity $\eta = 1.8$ kg$/$m$\cdot$s, while $g=9.81$ m$/$s$^2$. 
\begin{table*}
\begin{center} 
\begin{tabular}{|c|c|c|c|c|c|c|c|c|}
\hline
\hline
$P$ (mbar) & $T$ (K) & $u_{{\ast}{\mathrm{t}}}$ (m$/$s) & $z_{\mathrm{m}}$ (m) & $z_1$ (m) & $\alpha$ & $C_{\mathrm{d}}$ & $v_{\mathrm{s}}$ (m$/$s) & $q_{\mathrm{s}}$ (kg$/$m$\cdot$s) \\ \hline \hline
$5.0$ & $150$ & $1.804$ & $0.789$ & $0.011$ & $0.300$ & $3.744$ & $13.132$ & $0.152$ \\ \hline
$5.0$ & $200$ & $2.162$ & $1.154$ & $0.014$ & $0.227$ & $5.043$ & $18.017$ & $0.170$ \\ \hline 
$5.0$ & $250$ & $2.487$ & $1.543$ & $0.017$ & $0.184$ & $6.505$ & $22.957$ & $0.187$ \\ \hline
$7.5$ & $150$ & $1.449$ & $0.553$ & $0.009$ & $0.339$ & $3.331$ & $9.964$ & $0.127$ \\ \hline
$7.5$ & $200$ & $1.736$ & $0.810$ & $0.012$ & $0.257$ & $4.389$ & $13.617$ & $0.141$ \\ \hline
$7.5$ & $250$ & $1.996$ & $1.082$ & $0.015$ & $0.209$ & $5.567$ & $17.338$ & $0.154$ \\ \hline
$10.0$ & $150$ & $1.241$ & $0.431$ & $0.008$ & $0.371$ & $3.083$ & $8.205$ & $0.111$ \\ \hline
$10.0$ & $200$ & $1.486$ & $0.630$ & $0.011$ & $0.280$ & $4.001$ & $11.173$ & $0.123$ \\ \hline
$10.0$ & $250$ & $1.708$ & $0.841$ & $0.014$ & $0.228$ & $5.015$ & $14.210$ & $0.135$ \\ \hline \hline \hline \hline \hline \hline
$1000$ & $300$ & $0.218$ & $0.016$ & $0.004$ & $0.431$ & $2.747$ & $1.419$ & $0.009$ \\ \hline \hline \hline
\end{tabular}
\end{center}
\vspace{-0.5cm}
\caption{Main quantities controlling saltation on Mars under several values of pressure $P$ and temperature $T$, and a {\em{constant}} $u_{\ast}/u_{{\ast}{\mathrm{t}}} = 1.5$. The threshold shear velocity $u_{{\ast}{\mathrm{t}}}$, the mean saltation height $z_{\mathrm{m}}$, the drag coefficient $C_{\mathrm{d}}$, and the model variables $z_1$ and $\alpha$ depend on the atmospheric conditions, and have been calculated for a constant grain diameter $d=500$ ${\mu}$m and density ${\rho}_{\mathrm{grain}} = 3200$ kg$/$m$^3$, and with a dynamic viscosity obtained from the temperature (eq. ({\ref{eq:eta}})). The grain velocity $v_{\mathrm{s}}$ and the saturated flux $q_{\mathrm{s}}$ have been calculated with eqs. (\ref{eq:vs_2D}) and (\ref{eq:saturated_flux}), respectively. The corresponding values for terrestrial saltation are shown for comparison. On Earth, the value $u_{\ast} = 1.5 u_{{\ast}{\mathrm{t}}}$ means a shear velocity of $0.32$ m$/$s.} \label{tab:parameters_Mars}
\end{table*}

We see in table \ref{tab:parameters_Mars} that the values of sand flux on Mars are typically 10 times larger than on Earth. This is in agreement with the findings from wind tunnel simulations of saltation in a Martian environment by White \cite{White_1979}. We also see that Martian particles travel with higher average velocities, while the mean saltation height $z_{\mathrm{m}}$ on Mars is larger than on Earth, and may be over $1.0$m depending on the atmospheric conditions.

\begin{figure*}
\begin{center}
\includegraphics[width=0.75 \textwidth]{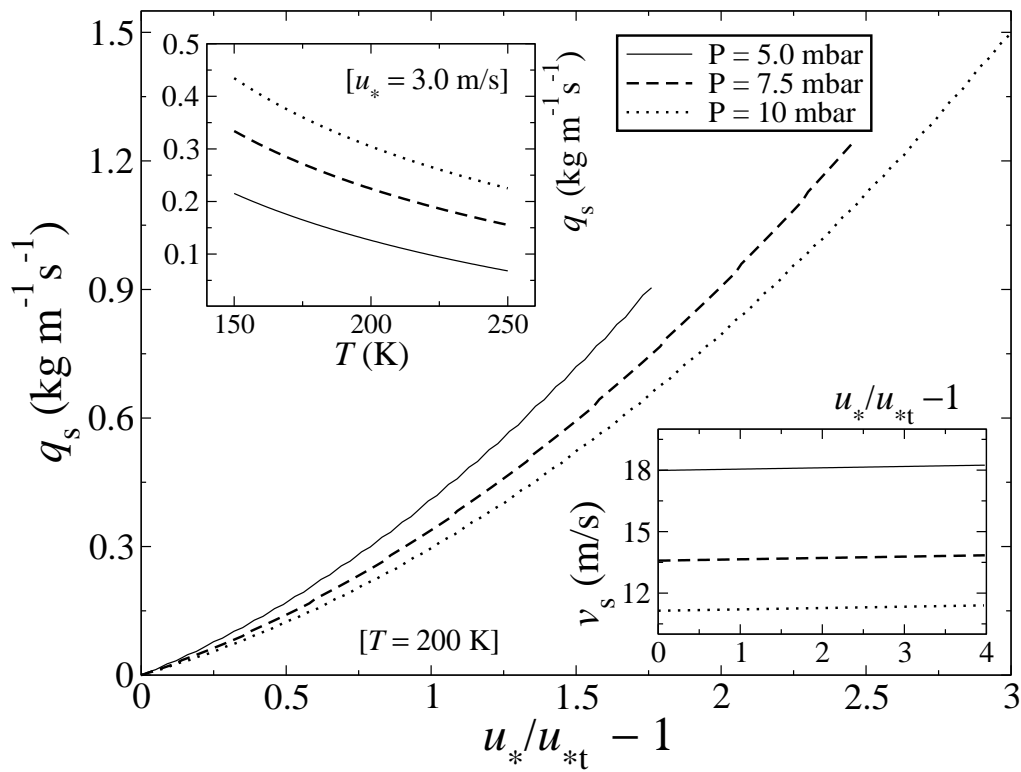}
\caption{Main plot: Saturated sand flux $q_{\mathrm{s}}$ as a function of the relative shear velocity $u_{\ast}/u_{{\ast}{\mathrm{t}}}-1$ for different values of atmospheric pressure --- and therefore different values of $u_{{\ast}{\mathrm{t}}}$ --- obtained with a temperature $T=200$ K. The lower inset on the right shows the corresponding values of the average grain velocity $v_{\mathrm{s}}$. In the upper inset on the left, we show the saturated flux for $u_{\ast} = 3.0$ m$/$s calculated for different values of temperature valid on Mars. }
\label{fig:q_u_star_Mars}
\end{center}
\end{figure*}

While the ratio $u_{\ast}/u_{{\ast}{\mathrm{t}}}$ in table \ref{tab:parameters_Mars} is constant, in the upper inset of fig. \ref{fig:q_u_star_Mars} we calculate $q_{\mathrm{s}}$ for a {\em{constant}} wind velocity $u_{\ast} = 3.0$ m$/$s using values of $P$ and $T$ within the range studied in table \ref{tab:parameters_Mars}. We see that the same wind friction speed transports more sand where $P$ is higher and $T$ is lower, which means lower ${\rho}_{\mathrm{fluid}}$ and $u_{{\ast}{\mathrm{t}}}$.

In the main plot and the lower inset of fig. \ref{fig:q_u_star_Mars} we show how the flux and the particle velocity at a given location on Mars depend on $u_{\ast}/u_{{\ast}{\mathrm{t}}}$. In the main plot we fix $T=200$ K and calculate the saturated flux $q_{\mathrm{s}}$ for different values of atmospheric pressure $P$ as function of $u_{\ast}/u_{{\ast}{\mathrm{t}}} - 1$. In the lower inset, the same calculations are made for the grain velocity $v_{\mathrm{s}}$. We see that the grain velocity in equilibrium is determined by the atmospheric conditions and has only a weak dependence on the friction speed $u_{\ast}$. The equilibrium velocity of the grains, $v_{\mathrm{s}}$, in fact scales with $u_{{\ast}{\mathrm{t}}}$. Equation (\ref{eq:velocity}) can be analytically solved in the simple case of the two-dimensional flow over a sand bed, where the gravitational term can be disregarded, which gives \cite{Duran_and_Herrmann_2006b}
\begin{equation}
v_{\mathrm{s}} = u_{\mathrm{eff}} - {v_{\mathrm{f}}^{\ast}}/{\sqrt{2{\alpha}}}. \label{eq:vs_2D}
\end{equation}
Because $u_{\mathrm{eff}}$ (eq. ({\ref{eq:u_eff}})) and $v_{\mathrm{f}}^{\ast}$ (eq. ({\ref{eq:v_f}})) both scale with $u_{{\ast}{\mathrm{t}}}$, $v_{\mathrm{s}}$ also does. 

In fig. \ref{fig:q_u_star_Mars} we see that for a given value of $u_{\ast}/u_{{\ast}{\mathrm{t}}}$, both the flux and the grain velocity are larger for lower atmospheric pressure $P$. This is because the shear velocity $u_{{\ast}{\mathrm{t}}}$ required for sand transport is higher for lower $P$, while $v_{\mathrm{s}}$ scales with $u_{{\ast}{\mathrm{t}}}$ and $q_{\mathrm{s}}$ scales with $u_{{\ast}{\mathrm{t}}}^2$ (eq. (\ref{eq:saturated_flux})). 

Table \ref{tab:velocity} shows $v_{\mathrm{s}}$ and $q_{\mathrm{s}}$ calculated for different values of $u_{\ast}/u_{{\ast}{\mathrm{t}}}$ on Mars and on Earth. Because the threshold wind friction speed on Mars is 10 times higher than on Earth, the average velocity of saltating grains on Mars is one order of magnitude higher than the velocity of Earth's grains. Again, $v_{\mathrm{s}}$ may have different values depending on the location on Mars, while $q_{\mathrm{s}}$ depends further on $u_{\ast}$.

\begin{table}[!t]
\begin{center} 
\begin{tabular}{|c|||c|c|||c|c|}
\hline
\hline
$u_{\ast}/u_{{\ast}{\mathrm{t}}}$ & $v_{\mathrm{s}}$ (m$/$s) & $v_{\mathrm{s}}$ (m$/$s) & $q_{\mathrm{s}}$ (m$/$s) & $q_{\mathrm{s}}$ (m$/$s)\\
$    $ & [Earth] & [Mars] & [Earth] & [Mars] \\ \hline \hline
$1.05$ & $1.367$ & $15.854$ & $0.0007$ & $0.0128$ \\ \hline
$1.10$ & $1.373$ & $15.857$ & $0.0015$ & $0.0262$ \\ \hline 
$1.25$ & $1.390$ & $15.867$ & $0.0040$ & $0.0703$ \\ \hline
$1.50$ & $1.419$ & $15.883$ & $0.0090$ & $0.1563$ \\ \hline
$1.70$ & $1.442$ & $15.896$ & $0.0139$ & $0.2366$ \\ \hline
$2.00$ & $1.447$ & $15.916$ & $0.0226$ & $0.3760$ \\ \hline \hline 
\end{tabular}
\end{center}
\caption{Average velocity ${v}_{\mathrm{s}}$ of saltating grains on Earth and on Mars as a function of the relative shear velocity $u_{\ast}/u_{{\ast}{\mathrm{t}}}$. Temperature $T=200$ K and pressure $P = 6.0$ mbar were used for Mars.} \label{tab:velocity}
\end{table}

In summary, using the atmospheric data provided by MGS Radio Science Team, we can calculate the quantities controlling saltation at a given location on Mars, for example, at a given dune field. From the ``Weather'' maps \cite{MGSRS}, we obtain the value of $P$ and $T$ characteristic of the area at which the dune field is located. Next, the density and viscosity are calculated from $P$ and $T$, using the ideal gas law and eq. (\ref{eq:eta}), while the model parameters are obtained, as exemplified in Table \ref{tab:parameters_Mars}, using the grain diameter $d=500$ ${\mu}$m of Martian sand dunes. 

The wind velocity $u_{\ast}$ in the dune fields on Mars is an unknown quantity. It must be determined from the calculations of dunes, as we will see in the next Section. 

Indeed, there is still one missing quantity for Mars which we need in order to solve the sand transport equations: ${\gamma}$, which appears in eq. (\ref{eq:saturation_length}). $\gamma$ is given by the product $r{\tilde{\gamma}}$, where $r$ (eq. (\ref{eq:r_definition})) is related to the saltation trajectories, and ${\tilde{\gamma}}$ (eq. (\ref{eq:splash})) gives the strength of the soil erosion. However, $r$ and ${\tilde{\gamma}}$ can not be calculated separately \cite{Sauermann_et_al_2001}. It is the quantity $\gamma$ (eqs. (\ref{eq:saturation_length}) and (\ref{eq:r_gamma})) that can be determined from comparison with measurements of the transient of flux saturation. The terrestrial value $\gamma = 0.2$ has been obtained by Sauermann {\em{et al.}} \cite{Sauermann_et_al_2001} from comparison of the saturation transient of the flux with experimental and numerical data \cite{Anderson_and_Haff_1991,McEwan_and_Willetts_1991,Butterfield_1993}, which are not available for Mars. Therefore, ${\gamma}$ is the parameter of the saltation model that remains to be determined for saltation transport on Mars. It will be obtained in the next Section, from the calculations of Martian barchan dunes.


\section{\label{sec:dunes}Dune formation on Mars}

\begin{figure}[htpb]
\begin{center}
\includegraphics[width=0.8 \columnwidth]{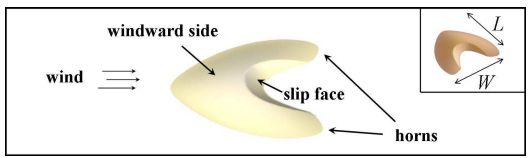}
\caption{Sketch of a barchan dune showing the windward side, horns and slip face. In the inset we see the definitions of dune width $W$ and length $L$.}
\label{fig:barchan_sketch}
\end{center}
\end{figure}

One very common type of dunes on Mars are {\em{barchans}}. They have one slip face and two horns, and propagate on bedrock under conditions of uni-directional wind (fig. \ref{fig:barchan_sketch}). Barchans are the simplest and best known dunes \cite{Bagnold_1941,Finkel_1959,Long_and_Sharp_1964,Hastenrath_1967,Lettau_and_Lettau_1969,Embabi_and_Ashour_1993,Hesp_and_Hastings_1998}. They are the subject of scientific and also environmental interest because of their high migration speed: on Earth, barchans $1-5$ m high may cover $30-100$ m in a year.

On Mars, barchan dunes occur on the floor of craters and on the north pole \cite{Bourke_et_al_2004}. Similarly to Earth's barchans, they form corridors and large dune fields, as in the Arkhangelsky crater (fig. \ref{fig:MOC_images}a). It appears surprising that intra-crater dunes on Mars look in general similar: they have mostly an elongated shape \cite{Bourke_et_al_2004}. 

The barchan dunes in the Arkhangelsky crater ($41.0^{\circ}$S, $25.0^{\circ}$W) are amongst the largest barchans on Mars. Further, there are good reasons to begin our study of Martian barchans with the Arkhangelsky dunes: they have a wide spectrum of dune sizes; have not been significantly altered by secondary winds, and do not appear joined at their horns forming chains of barchanoids. For example, such features are observed in the dunes at Kaiser crater and Proctor crater \cite{Fenton_et_al_2003}.

Let us try to reproduce the shape of the Arkhangelsky barchans with the dune model using parameters of the present atmosphere of Mars. 

The atmospheric pressure $P$ and temperature $T$ near the Arkhangelsky crater are, respectively, $5.5$ mbar and $210$ K \cite{MGSRS}. These values yield a local Martian atmospheric density ${\rho}_{\mathrm{fluid}} = 0.014$ kg$/$m$^3$, and a fluid viscosity $\eta \approx 1.06\,{\mbox{kg}}/$m$\cdot$s. Using the mean grain diameter $d=500$ ${\mu}$m, grain density ${\rho}_{\mathrm{grain}} = 3200$ kg$/$m$^3$, and gravity $3.71$ m$/$s$^2$, it follows that the threshold wind friction speed for saltation in the Arkhangelsky crater is $u_{{\ast}{\mathrm{t}}} = 2.12$ m$/$s. In this manner, all parameters of the saltation model, but $\gamma$ (eq. ({\ref{eq:r_gamma}})), are determined using the equations presented in the previous Section.


\subsection{The shape of the Arkhangelsky barchans}

In the calculations of barchans, a constant upwind shear velocity $u_{\ast} > u_{{\ast}{\mathrm{t}}}$ and a small influx $q_{\mathrm{in}}/q_{\mathrm{s}}$ (the interdune flux) are imposed in $x$ (downwind) direction at the inlet, starting with a Gaussian hill having the volume similar to the dune we want to reproduce. The hill evolves in time until displaying the barchan {\em{shape}}, i.e. linear relations between length $L$, width $W$ and height $H$ \cite{Kroy_et_al_2002,Parteli_et_al_2007}.

Measured values of shear velocity of Martian winds are mostly between $0.4$ and $0.6$ m$/$s \cite{Sutton_et_al_1978}, which are much lower values than the Martian threshold for saltation ($\approx 2.0$ m$/$s). Indeed, many authors estimated that the shear velocity $u_{\ast}$ of Martian sand-moving winds, which occur within gusts of extreme dust storms, may reach maximum values between $2.2$ and $4.0$ m$/$s \cite{Arvidson_et_al_1983,Moore_1985}. Again, very unprobably $u_{\ast}$ must achieve values of the order of $4.0$ m$/$s on Mars \cite{Sullivan_et_al_2005}.

On the other hand, we know from experience with terrestrial dune fields that the flux in areas between dunes is normally small, between $10$ and $40\%$ of the maximum flux $q_{\mathrm{s}}$ \cite{Fryberger_et_al_1984}. Moreover, as shown from calculations in previous work, the shape of a barchan dune of given size depends in an important manner on the interdune flux only for values of $q_{\mathrm{in}}/q_{\mathrm{s}}$ above this range \cite{Parteli_et_al_2007}. 

On the basis of the observations obove, we try to reproduce, first, the shape of {\em{one}} Arkhangelsky barchan, which has width $W \approx 650$ m, using $q_{\mathrm{in}}/q_{\mathrm{s}} = 20\%$. Furthermore, we take values of wind friction speed in the maximum range between $2.0$ and $4.0$ m$/$s, which gives $u_{\ast}/u_{{\ast}{\mathrm{t}}}$ approximately between $1.0$ and $2.0$ in the Arkhangelsky crater. Now the quantity $\gamma$ remains (eq. ({\ref{eq:r_gamma}})), which could not be estimated for Mars. As a first guess, we take the terrestrial value $\gamma = 0.2$ and investigate whether the Arkhangelsky dune can be obtained with $u_{\ast}$ in the above mentioned range.

\begin{figure}
  \begin{center}
   \includegraphics*[width=1.0 \columnwidth]{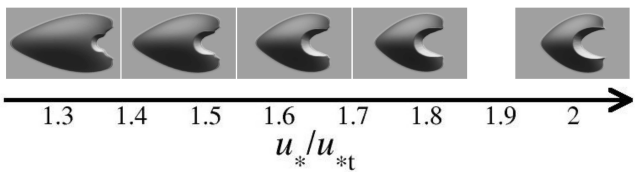}
 \caption{Barchan dunes of width $W=650$ m calculated using parameters for Mars, with $\gamma = 10\,{\gamma}_{\mathrm{Earth}}$, and different values of wind shear velocity $u_{\ast}/u_{{\ast}{\mathrm{t}}}$.}
     \label{fig:simArkhangelsky}
 \end{center}
 \end{figure}

We obtained a surprising result: if we take the same $\gamma=0.2$ as on Earth, the Gaussian hill does not evolve into a barchan: it simply does not develop a slip face but a {\em{dome}} is obtained. However, if we take ${\gamma}_{\mathrm{Mars}}$ on Mars of the order of $10$ times the terrestrial value $\gamma_{\mathrm{Earth}} = 0.2$, then barchan dunes with shape similar to the Arkhangelsky barchan can be obtained, as we see in fig. \ref{fig:simArkhangelsky}: the elongated shape characteristic of the Arkhangelsky dunes is a result of low values of shear velocity,  with $u_{\ast}/u_{{\ast}{\mathrm{t}}}$ smaller than $1.5$. This means values of $u_{\ast}$ up to $3.0$ m$/$s. For a constant value of $\gamma = 2.0$, the dune shape deviates from the Arkhangelsky barchans for increasing values of $u_{\ast}$. Thus, the shear velocity in the Arkhangelsky crater must be close to the threshold for saltation transport. This explains the elongated shape of intra-crater barchan dunes on Mars.

\subsection{Entrainment of saltating grains on Mars}

The wind shear velocity $u_{\ast} \approx 3.0$ m$/$s estimated for the Arkhangelsky crater is well within predicted maximum values of $u_{\ast}$ on Mars. But why should the Martian $\gamma$ (eq. (\ref{eq:r_gamma})) be ten times larger than on Earth? 

The quantity $r = {\gamma}/{\tilde{\gamma}}$ (eq. (\ref{eq:r_definition})) on Mars should not differ much from the Earth's value. This is because the ejection velocity of splashed grains is proportional to the velocity of the average impacting grains \cite{Anderson_and_Haff_1988}, which in turn scales with the average saltation velocity $v_{\mathrm{s}}$. In this manner, we must understand why the Martian entrainment rate, $\tilde{\gamma}$, differs from the one on Earth. This quantity determines the intensity of the grain-bed collisions, the modelling of which is beyond the scope of this work \cite{Sauermann_et_al_2001}.

However, Anderson and Haff \cite{Anderson_and_Haff_1988} showed that the number of splashed grains is proportional to the velocity $v_{\mathrm{imp}}$ of the impacting grains. Let us rescale $v_{\mathrm{imp}}$ with $v_{\mathrm{esc}} = {\sqrt{gd}}$, which is the velocity necessary to escape from the sand bed \cite{Andreotti_et_al_2002}. This velocity has value approximately $4.5$ cm$/$s, both on Mars and on Earth. Further, $v_{\mathrm{imp}}$ scales with the mean grain velocity, $v_{\mathrm{s}}$. This leads to the following scaling relation for the entrainment rate of saltating grains:
\begin{equation}
\tilde{\gamma} \propto v_{\mathrm{s}}/{\sqrt{gd}}.  \label{eq:gamma_vs}
\end{equation} 
Typical values of the average velocity of saltating grains on Mars are shown in fig. \ref{fig:q_u_star_Mars} and in Table \ref{tab:velocity} as function of the relative wind friction speed $u_{\ast}/u_{{\ast}{\mathrm{t}}}$. We see that the grain velocity on Mars is one order of magnitude larger than on Earth: $v_{\mathrm{s}}$ scales with $u_{{\ast}{\mathrm{t}}}$ and has only a very weak dependence on $u_{\ast}$ which we neglect. In this manner, we can write ${\tilde{\gamma}} \propto u_{{\ast}{\mathrm{t}}}/{\sqrt{gd}}$. Since we know that $\gamma = 0.2$ on Earth, where $g=9.81$ m$/$s$^2$, $d=250$ ${\mu}$m and $u_{{\ast}{\mathrm{t}}} = 0.218$ m$/$s, we obtain
\begin{equation}
\gamma = 0.045{\frac{u_{{\ast}{\mathrm{t}}}}{\sqrt{gd}}}. \label{eq:gamma_equation}
\end{equation}
Equation (\ref{eq:gamma_equation}) gives $\gamma \approx 2.24$ in the Arkhangelsky crater, which is in fact one order of magnitude higher than the Earth's value, as obtained in a different way from the calculations in fig. \ref{fig:simArkhangelsky}. 

Summarizing, we found that the entrainment rate of grains into saltation on Mars is ten times higher than on Earth. This is explained by the Martian larger splash events, which are consequence of the higher average velocity of saltating grains on Mars. 

What is the consequence of a ten times higher entrainment rate on Mars? Because the saturation length of the sand flux depends on the rate at which grains enter saltation, the larger splash events on Mars have a crucial implication for the formation of sand dunes on the red planet.

While on one hand the lower Martian atmospheric density ${\rho}_{\mathrm{fluid}}$ and gravity $g$ result in longer grain trajectories than on Earth \cite{White_1979}, the saturation transient of the flux on Mars is shortened by the faster increase in the population of saltating grains. This is because the wind strength is reduced more rapidly the faster the grains are launched into saltation after splash (``feedback effect'' \cite{Owen_1964}). In fact, the characteristic length of flux saturation, ${\ell}_{\mathrm{s}}$ (eq. (\ref{eq:saturation_length})), scales with the average saltation length, $\ell$. However, ${\ell}_{\mathrm{s}}$ is, furthermore, proportional to $1/{\tilde{\gamma}}$. 

\begin{figure}[!t]
\begin{center}
\includegraphics[width=1.0 \columnwidth]{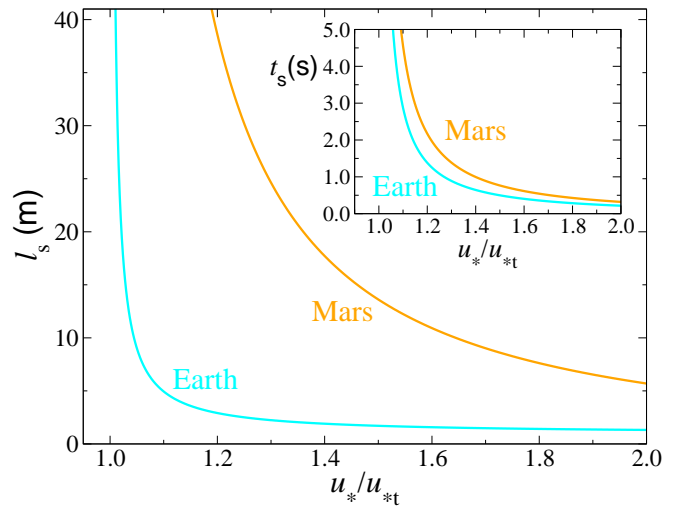}
\caption{Main plot: characteristic length of flux saturation, ${\ell}_{\mathrm{s}}$, calculated with eq. (\ref{eq:saturation_length}), as function of $u_{\ast}/u_{{\ast}{\mathrm{t}}}$. The inset shows the characteristic time $t_{\mathrm{s}} = {\ell}_{\mathrm{s}}/v_{\mathrm{s}}$ as function of  $u_{\ast}/u_{{\ast}{\mathrm{t}}}$, where $v_{\mathrm{s}}$ is the average grain velocity (eq. (\ref{eq:vs_2D})).}
\label{fig:saturation_time}
\end{center}
\end{figure}

The characteristic length of flux saturation, ${\ell}_{\mathrm{s}}$, is calculated in the main plot of fig. \ref{fig:saturation_time} using parameters for Earth and for the Arkhangelsky crater on Mars, with $u_{\ast}/u_{{\ast}{\mathrm{t}}}$ in the range between $1.0$ and $2.0$. In this figure, the Martian ${\ell}_{\mathrm{s}}$ has been calculated using eq. (\ref{eq:saturation_length}) with $\gamma$ given by eq. (\ref{eq:gamma_equation}). In the inset of fig. \ref{fig:saturation_time}, we have calculated the characteristic time of flux saturation, $t_{\mathrm{s}} = {\ell}_{\mathrm{s}}/{v}_{\mathrm{s}}$, for different values of $u_{\ast}/u_{{\ast}{\mathrm{t}}}$. 

It is remarkable that although the Martian and terrestrial values of ${\ell}_{\mathrm{s}}$ differ by a factor of 10, $t_{\mathrm{s}}$ on Mars is nearly the same as the terrestrial one for a given $u_{\ast}/u_{{\ast}{\mathrm{t}}}$. This is because the average velocity of saltating grains, $v_{\mathrm{s}}$, is one order of magnitude higher on Mars, as shown in Section \ref{sec:saltation}. 

\begin{figure}[!t]
\begin{center}
\includegraphics[width=1.0 \columnwidth]{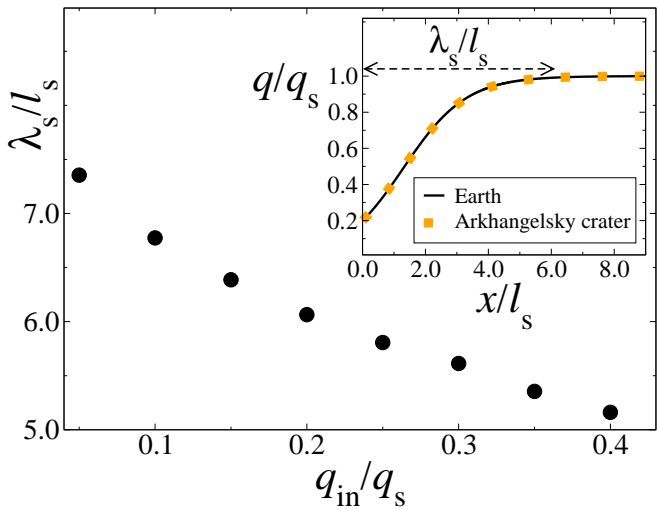}
\caption{Sand flux calculated over a flat sand bed on Mars and on Earth. Main plot: distance of flux saturation, ${\lambda}_{\mathrm{s}}$, normalized by the characteristic length ${\ell}_{\mathrm{s}}$, as function of the influx $q_{\mathrm{in}}/q_{\mathrm{s}}$ at the inlet. Inset: evolution of the normalized flux $q/q_{\mathrm{s}}$ with downwind distance $x/{\ell}_{\mathrm{s}}$ for $q_{\mathrm{in}}/q_{\mathrm{s}} = 0.20$.}
\label{fig:lambda_s}
\end{center}
\end{figure}

As an example, we calculate the sand flux $q$ (eq. (\ref{eq:differential_q})), over a flat sand bed submitted to an unidirectional wind of constant strength, using parameters for Earth and for the Arkhangelsky crater on Mars. A constant influx $q_{\mathrm{in}}/q_{\mathrm{s}}$ is set at the inlet. The evolution of the normalized sand flux $q/q_{\mathrm{s}}$ with the downwind distance $x/{\ell}_{\mathrm{s}}$ calculated using parameters for Earth (line) and for the Arkhangelsky crater (symbols) is shown in the inset of fig. \ref{fig:lambda_s} with $q_{\mathrm{in}}/q_{\mathrm{s}} = 0.2$. In the main plot of fig. \ref{fig:lambda_s}, we see that it takes a distance ${\lambda}_{\mathrm{s}}$ of about $6$ times ${\ell}_{\mathrm{s}}$, from the edge of the sand bed, for the sand flux to achieve $99\%$ of its saturated value $q_{\mathrm{s}}$, using realistic values of $q_{\mathrm{in}}/q_{\mathrm{s}}$ between $0.1$ and $0.4$. 

If we take, for instance, $u_{\ast}/u_{{\ast}{\mathrm{t}}} = 1.50$, then we obtain, with eq. (\ref{eq:saturation_length}), ${\ell}_{\mathrm{s}} = 0.71$ and $13.62$ m on Earth, respectively in the Arkhangelsky crater. On the basis of fig. \ref{fig:lambda_s}, this leads to ${\lambda}_{\mathrm{s}} \approx 4.3$ m on Earth, while the Martian ${\lambda}_{\mathrm{s}}$ is approximately $81$ m. Furthermore, flux saturation is reached within approximately $6\,t_{\mathrm{s}}$, i.e. within $3.0$ s on Earth and $4.6$ s on Mars. However, if we had taken the terrestrial $\gamma = 0.2$ for Mars, then the values of Martian flux transient length and time obtained would be of the order of 100, respectively 10 times larger than the terrestrial ones.

The larger value of ${\gamma}$ on Mars shortens the characteristic distance of flux saturation (eq. (\ref{eq:saturation_length})) by one order of magnitude. However, since dunes cannot be smaller than the saturation length, this means that the scale of dunes that is predicted from the scaling of the flux saturation distance with the average saltation length, $\ell$ \cite{Kroy_et_al_2005}, is reduced by a factor of 10. In conclusion, the Martian larger splash is the missing link to understand the size of Martian dunes formed by the thin atmosphere of the red planet.

\subsection{Wind speed and migration velocity of barchans}

In the calculations of fig. \ref{fig:simArkhangelsky}, we could estimate the shear velocity $u_{\ast}/u_{{\ast}{\mathrm{t}}}$ from comparison with the shape of one selected barchan in the field, taking ${\gamma}_{\mathrm{Mars}} = 10\,{\gamma}_{\mathrm{Earth}}$ and assuming a given value of the interdune flux $q_{\mathrm{in}}/q_{\mathrm{s}}$ in the Arkhangelsky crater. However, since $\gamma$ can be now {\em{calculated}} for different atmospheric conditions, we can solve the model equations to find the values of $u_{\ast}/u_{{\ast}{\mathrm{t}}}$ and $q_{\mathrm{in}}/q_{\mathrm{s}}$ at a given dune field on Mars. Indeed, both field variables can be obtained from comparison with the {\em{minimal dune}} \cite{Parteli_et_al_2007}, as summarized in the next paragraph.

In order to develop the slip face characteristic of barchans, sand hills must reach a minimum size, below which they are called {\em{domes}}. As shown in a previous work, the minimal dune width $W_{\mathrm{min}}$ is around $13{\ell}_{\mathrm{s}}$, and is approximately independent of the interdune flux, $q_{\mathrm{in}}/q_{\mathrm{s}}$ \cite{Parteli_et_al_2007}. In this manner, $W_{\mathrm{min}}$ yields, through eq. (\ref{eq:saturation_length}), the value of $u_{\ast}/u_{{\ast}{\mathrm{t}}}$ at a given dune field. Moreover, once $u_{\ast}/u_{{\ast}{\mathrm{t}}}$ is determined, the value of $q_{\mathrm{in}}/q_{\mathrm{s}}$ can be obtained from the shape of the minimal dune: the eccentricity $L_{\mathrm{min}}/W_{\mathrm{min}}$ decreases approximately linearly with  $q_{\mathrm{in}}/q_{\mathrm{s}}$ \cite{Parteli_et_al_2007}.

\begin{figure}[!t]
\begin{center}
\includegraphics[width=0.85 \columnwidth]{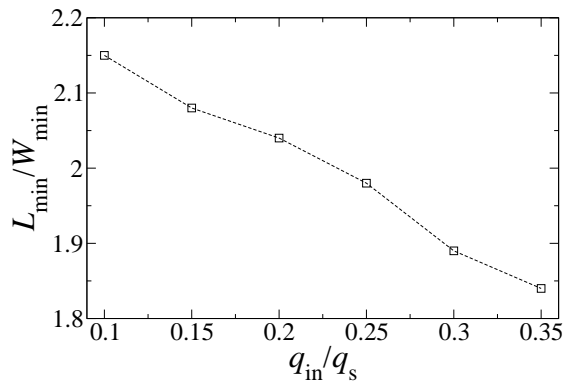}
\caption{Eccentricity of the minimal dune as function of the average interdune flux, $q_{\mathrm{in}}/q_{\mathrm{s}}$, calculated using $u_{\ast}/u_{{\ast}{\mathrm{t}}} \approx 1.45$, $P=5.5$ mbar and $T=210$ K. We see that the eccentricity $L_{\mathrm{min}}/W_{\mathrm{min}} \approx 2.0$ of the domes in the Arkhangelsky Crater is reproduced with an interdune flux of $25\%$ of the saturated flux.}
\label{fig:Calculations_Arkhangelsky}
\end{center}
\end{figure}

In the Arkhangelsky crater on Mars (fig. \ref{fig:MOC_images}a), the minimal dune is indicated by two domes which have width $W_{\mathrm{min}} \approx 200$ m and length $L_{\mathrm{min}} \approx 400$ m. From $W_{\mathrm{min}} = 200$ m, we obtain ${\ell}_{\mathrm{s}} \approx 15.5$ m, which gives $u_{\ast}/u_{{\ast}{\mathrm{t}}} \approx 1.45$ or $u_{\ast} \approx 3.07$ m$/$s in the Arkhangelsky crater. This is essentially the same result obtained previously from comparison with the elongated shape (fig. \ref{fig:simArkhangelsky}). Next, using this shear velocity, we calculate the eccentricity $L_{\mathrm{min}}/W_{\mathrm{min}}$ of the minimal ``Arkhangelsky'' dune as function of $q_{\mathrm{in}}/q_{\mathrm{s}}$. We see in fig. \ref{fig:Calculations_Arkhangelsky} that the ratio $L_{\mathrm{min}}/W_{\mathrm{min}} \approx 2.0$ is obtained with an average interdune flux $q_{\mathrm{in}} \approx 25\%$ of the saturated flux $q_{\mathrm{s}}$. Again, this value is nearly the same interdune flux assumed in the calculations of fig. \ref{fig:simArkhangelsky}.

\begin{figure*}[!t]
  \begin{center}
   \includegraphics*[width=0.6 \textwidth]{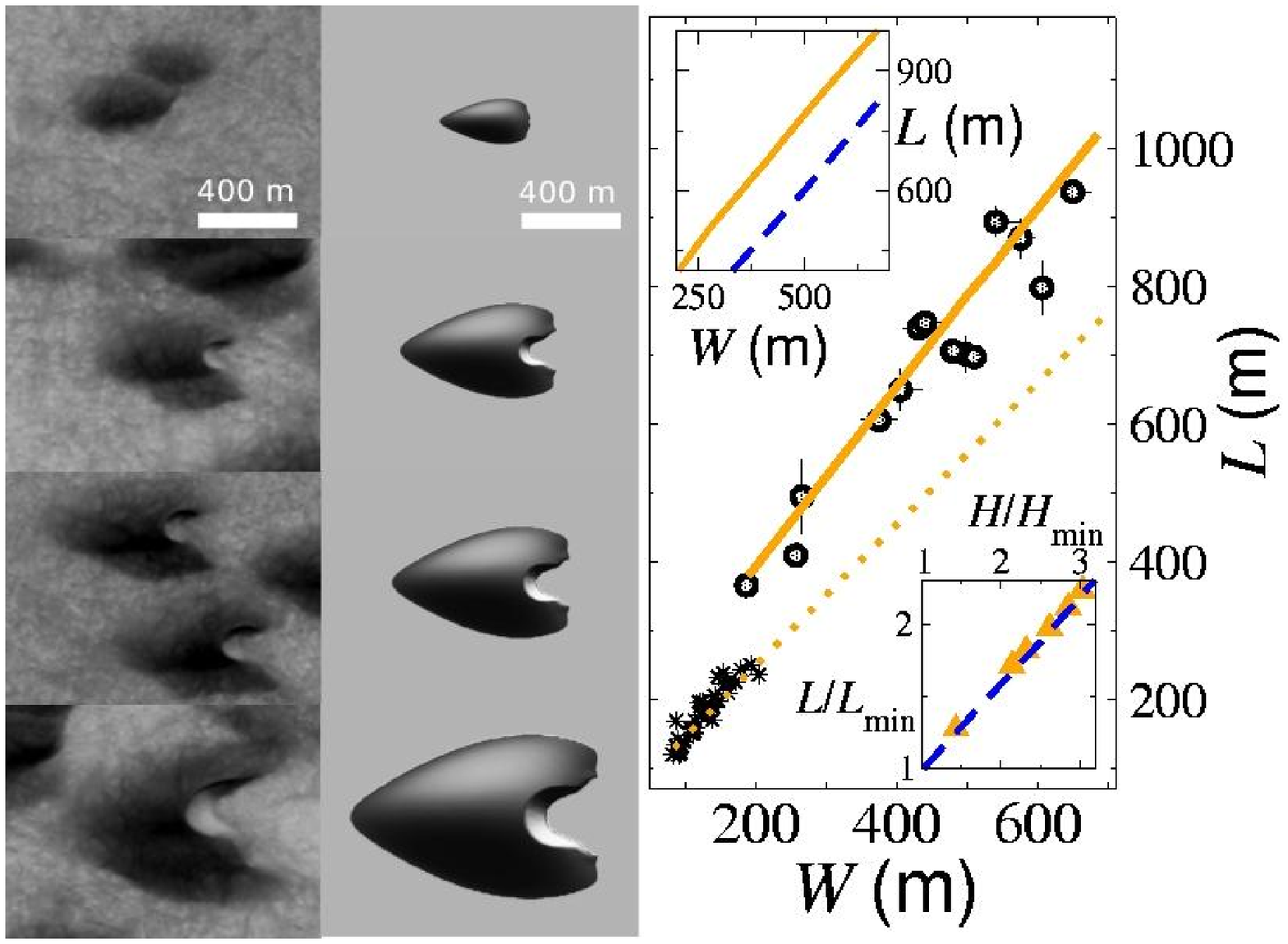}
 \caption{Barchans in the Arkhangelsky crater, $41.0^{\circ}$S, $25.0^{\circ}$W on Mars: Mars Orbiter Camera (MOC) images on the left (image courtesy of NASA$/$JPL$/$MSSS) and calculated dunes on the right. Main plot: $L$ vs $W$ of the Arkhangelsky (circles) and north polar barchans at $77.6^{\circ}$N, $103.6^{\circ}$W (stars). Calculations of Arkhangelsky (north polar) dunes are represented by the continuous (dotted) line, obtained with $u_{\ast}/u_{{\ast}{\mathrm{t}}}=1.45$ ($1.80$). The dashed line in the upper inset corresponds to terrestrial dunes obtained with $u_{\ast}/u_{{\ast}{\mathrm{t}}} = 1.45$. In the lower inset, we see $L/L_{\mathrm{min}}$ vs $H/H_{\mathrm{min}}$ from calculations of the Arkhangelsky (triangles) and terrestrial dunes (dashed line).}
     \label{fig:Mars_Arkhangelsky}
 \end{center}
 \end{figure*}

Figure \ref{fig:Mars_Arkhangelsky} shows the results obtained using $u_{\ast} = 1.45\,u_{{\ast}{\mathrm{t}}}$ and $q_{\mathrm{in}}/q_{\mathrm{s}} = 0.25$. In this figure, we show four Arkhangelsky dunes of different sizes next to dunes calculated with the model. Further, the main plot in fig. \ref{fig:Mars_Arkhangelsky} shows the length $L$ as function of width $W$ of the Arkhangelsky dunes (circles) and of the dunes obtained in calculations (full line). We see that the values of $u_{\ast}/u_{{\ast}{\mathrm{t}}}$ and $q_{\mathrm{in}}/q_{\mathrm{s}}$ obtained for the Arkhangelsky crater on Mars not only reproduce the minimal dune but also describe well the dependence of the shape on the dune size. 

In this manner, substituting eq. (\ref{eq:gamma_equation}) into eq. (\ref{eq:saturation_length}), we have obtained a closed set of sand transport equations that can be solved for different atmospheric conditions. Furthermore, using the model equations, the value of wind friction speed $u_{\ast}/u_{{\ast}{\mathrm{t}}}$ and interdune flux $q_{\mathrm{in}}/q_{\mathrm{s}}$ in a given dune field on Mars can be determined from comparison with the shape and the size of the minimal dune, on the basis of the results presented in Ref. \cite{Parteli_et_al_2007}.

\begin{figure}
  \begin{center}
   \includegraphics*[width=0.8 \columnwidth]{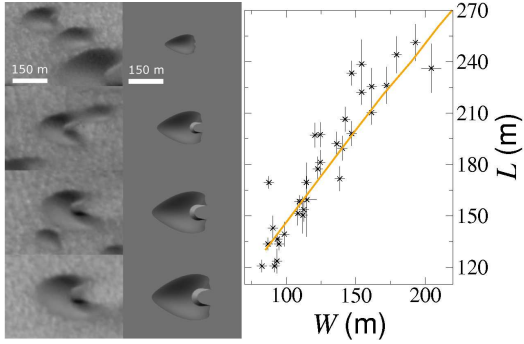}	
 \caption{Calculations of Martian north polar barchans near $77.6^{\circ}$N, $103.6^{\circ}$W (fig. \ref{fig:MOC_images}b). We see MOC images of dunes of different sizes on the left, and on the right we see dunes calculated using $P=8.0$ mbar, $T=190$ K, $u_{\ast}=2.92$ m$/$s and $q_{\mathrm{in}}/q_{\mathrm{s}} = 0.29$. The plot shows $L$ vs $W$ for the real dunes (stars) and for the calculated ones (dotted line), as also displayed in the main plot of fig. \ref{fig:Mars_Arkhangelsky} for comparison with the Arkhangelsky dunes.}
     \label{fig:north_polar_dunes}
 \end{center}
 \end{figure}

Let us study a second Martian barchan field which is near the north pole (fig. \ref{fig:MOC_images}b), and where $W_{\mathrm{min}} \approx 80$ m and $L_{\mathrm{min}} \approx 130$ m. At the location of the field, $P=8.0$ mbar and $T=190$ K \cite{MGSRS}, and thus $u_{{\ast}{\mathrm{t}}} \approx 1.62$ m$/$s. From the minimal dune width $W_{\mathrm{min}} = 80$ m, we obtain $u_{\ast}/u_{{\ast}{\mathrm{t}}} \approx 1.8$ or $u_{\ast} = 2.92$ m$/$s using eq. (\ref{eq:saturation_length}). Next, in the same way as done for the Arkhangelsky dunes, we calculate the eccentricity as function of the interdune flux, $q_{\mathrm{in}}/q_{\mathrm{s}}$, and we find that the value $L_{\mathrm{min}}/W_{\mathrm{min}} \approx 1.6$ is reproduced with $q_{\mathrm{in}}/q_{\mathrm{s}} = 0.30$. In fig. \ref{fig:north_polar_dunes}, we see that the behaviour $L$ against $W$ of the barchans in this field (stars) is well captured by the model (full line). Furthermore, these curves are also shown in the main plot of fig. \ref{fig:Mars_Arkhangelsky} (real and calculated north polar dunes are represented by stars, respectively, by the dotted line) for comparison with the Arkhangelsky dunes.

It is interesting that the $u_{\ast}$ obtained for the north polar field is very similar to that in the Arkhangelsky crater, although $u_{{\ast}{\mathrm{t}}}$ is lower in the north polar field due to the higher ${\rho}_{\mathrm{fluid}}$ (table \ref{tab:parameters_fields}). 
\begin{table}
\begin{center} 
\begin{tabular}{|c|c|c|c|c|}
\hline
Barchan field & ${\rho}_{\mathrm{fluid}}$ (kg$/$m$^3$) & $u_{{\ast}{\mathrm{t}}}$ (m$/$s) & $v_{\mathrm{g}}$ (m$/$s) & $\gamma$  \\ \hline \hline
Arkhangelsky & $0.014$ & $2.12$ & $17.8$ & $2.24$ \\ \hline
$77.6^{\circ}$N, $103.6^{\circ}$W & $0.022$ & $1.62$ & $12.3$ & $1.71$ \\ \hline
Earth & $1.225$ & $0.22$ & $1.5$ & $0.20$  \\ \hline 
\end{tabular}
\end{center}
\vspace{-0.5cm}
\caption{Main quantities controlling saltation on Mars and on Earth.} \label{tab:parameters_fields}
\end{table}

\begin{figure}
  \begin{center}
   \includegraphics*[width=1.0 \columnwidth]{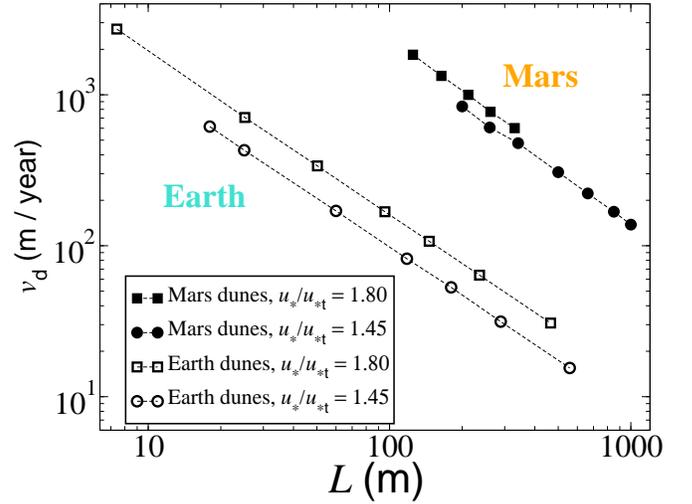}	
 \caption{Dune velocity $v_{\mathrm{d}}$ as function of dune length $L$. We see that Mars dunes (filled symbols) move typically ten times faster than Earth dunes (empty symbols) of same $L$, obtained with similar values of $u_{\ast}/u_{{\ast}{\mathrm{t}}}$ as on Mars.}
     \label{fig:dune_velocity}
 \end{center}
 \end{figure}

The values of $u_{\ast}/u_{{\ast}{\mathrm{t}}}$ obtained for Mars are within the range of the ones measured in terrestrial barchan fields \cite{Fryberger_and_Dean_1979,Embabi_and_Ashour_1993,Sauermann_et_al_2003}. Indeed, we see in fig. \ref{fig:dune_velocity} that, for the same value of $u_{\ast}/u_{{\ast}{\mathrm{t}}}$, Martian barchans would move ten times faster than those on Earth.

However, winds on Mars are only seldom above the threshold for saltation \cite{Sutton_et_al_1978,Arvidson_et_al_1983,Moore_1985}. As reported from observations of Mars missions, saltation transport on Mars occurs during a few seconds in time intervals of several years. If, for example, winds on Mars achieve $u_{\ast}\approx 3.0$ m$/$s during 40 s every 5 years \cite{Arvidson_et_al_1983}, then, from fig. \ref{fig:dune_velocity}, we see that a Martian barchan of length 200 m would need $[5$ years$]$ $\cdot 10^{-3} \cdot (3600 \cdot 24 \cdot 365)/40 \approx 4,000$ years to move $1.0$ m. This result explains why spacecrafts orbiting Mars never revealed any movement of Martian barchan dunes.

\subsection{Martian bimodal sand dunes}

Because linear and star dunes like the ones found on Earth were almost not observed in the first images of Mars taken by Mariner 9 and Viking orbiters, it has been suggested that non-unimodal wind regimes should be very rare on Mars \cite{Lee_and_Thomas_1995}. However, the Mars Global Surveyor MOC Camera has, more recently, imaged a high diversity of dune shapes that had been never observed in images of previous missions.

On bedrock and in areas of low sand availability, there appear many exotic and up to now unexplained dune forms where barchans should occur if the wind regime were uni-directional. Dunes as those in figs. \ref{fig:MOC_images}c$-$e cannot appear in areas of uni-directional winds, for in this case barchans should be formed. Indeed, it is possible to recognize in the images of figs. \ref{fig:MOC_images}c$-$e that the dominant winds define a resultant direction.

\begin{figure}
\begin{center}
\includegraphics*[width=1.0\columnwidth]{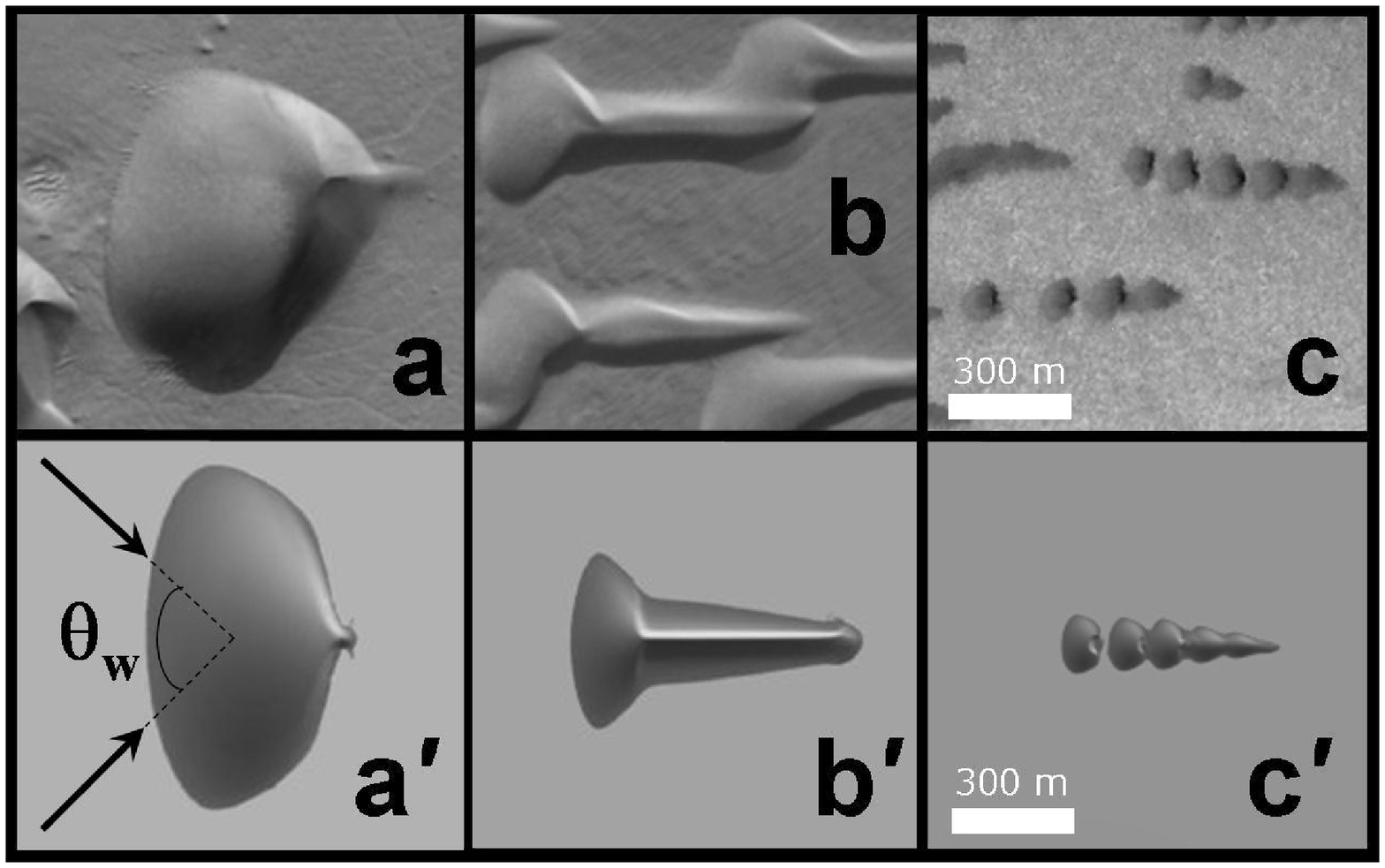}
\caption{MOC images on top (courtesy of NASA/JPL/MSSS) and calculations obtained with bimodal wind regimes on bottom. A sketch showing the definition of the angle ${\theta}_{\mathrm{w}}$ of the wind direction (arrows) is shown in a$^{\prime}$. The wind changes its direction with frequency $1/T_{\mathrm{w}}$. We chose $T_{\mathrm{w}} =$ $2.9$ days, $5.8$ days and $0.7$ day to obtain dunes in a$^{\prime}$, b$^{\prime}$ and $c^{\prime}$, respectively. The dune in a$^{\prime}$ has been obtained with ${\theta}_{\mathrm{w}} = 100^{\circ}$ and the dune in b$^{\prime}$ with ${\theta}_{\mathrm{w}}=140^{\circ}$. In c$^{\prime}$, a linear dune obtained with ${\theta}_{\mathrm{w}} = 120^{\circ}$ decays into barchans after the angle ${\theta}_{\mathrm{w}}$ is reduced to $80^{\circ}$. Dune in b$^{\prime}$ elongates to the right with time, which is not observed for the dune in a$^{\prime}$. Dune in c$^{\prime}$ decays further until only a string of rounded barchans remains.}
\label{fig:Mars_bimodal_dunes}
\end{center}
\end{figure}

We found that the dune shapes in figs. \ref{fig:MOC_images}c$-$e can be obtained with a {\em{bimodal}} wind regime. In our calculations, the wind {\em{alternates}} its direction periodically with frequency $1/T_{\mathrm{w}}$ forming an angle ${\theta}_{\mathrm{w}}$ as sketched in fig. \ref{fig:Mars_bimodal_dunes}a$^{\prime}$. In both directions the strength is the same, namely $u_{\ast} = 3.0$ m$/$s, as found from the calculations of barchan dunes. In this manner, the value of $u_{\ast}/u_{{\ast}{\mathrm{t}}}$ is particular to each field, since $u_{{\ast}{\mathrm{t}}}$ depends on the field location (table \ref{tab:parameters_bimodal_dunes}). 

To simulate the change of wind direction, we rotate the field by an angle ${\theta}_{\mathrm{w}}$, keeping the wind direction constant. In the calculations, thus, the separation bubble adapts to the wind direction after rotation of the field. We use open boundaries as in the calculations of barchan dunes. Initial condition is a Gaussian hill as before, whose volume is taken according to the volume of the dune. 

The angle ${\theta}_{\mathrm{w}}$ between the wind directions determines which of the different forms in fig. \ref{fig:Mars_bimodal_dunes} is obtained. We found that a barchan moving in the resulting wind direction is always obtained if ${\theta}_{\mathrm{w}} < 90^{\circ}$. If ${\theta}_{\mathrm{w}} \approx 100^{\circ}$, then the dune shape in fig. \ref{fig:Mars_bimodal_dunes}a$^{\prime}$ is achieved. And for ${\theta}_{\mathrm{w}}$ of the order of $120^{\circ}$ or larger, elongated dune forms as those in fig. \ref{fig:Mars_bimodal_dunes}b$^{\prime}$ are obtained, which elongate in time. As ${\theta}_{\mathrm{w}} \longrightarrow 180^{\circ}$, a dune form of alternating slip face position appears.

\begin{table}[htpb]
\begin{center} 
\begin{tabular}{|c|c|c|c|}
\hline
Field & location & ${\rho}_{\mathrm{fluid}}$ (kg$/$m$^3$) & $u_{{\ast}{\mathrm{t}}}$ (m$/$s) \\ \hline \hline
fig. \ref{fig:Mars_bimodal_dunes}a & $48.6^{\circ}$S, $25.5^{\circ}$W & $0.017$ & $1.89$ \\ \hline 
fig. \ref{fig:Mars_bimodal_dunes}b & $49.6^{\circ}$S, $352.9^{\circ}$W & $0.014$ & $2.06$  \\ \hline 
fig. \ref{fig:Mars_bimodal_dunes}c & $76.4^{\circ}$N, $272.9^{\circ}$W & $0.03$ & $1.35$ \\ \hline 
\end{tabular}
\end{center}
\vspace{-0.5cm}
\caption{For each dune field in fig. \ref{fig:Mars_bimodal_dunes}, the fluid density ${\rho}_{\mathrm{fluid}}$ and the threshold $u_{{\ast}{\mathrm{t}}}$ are calculated from the local pressure and temperature which are taken from the MGS Radio Science data (MGSRS 2006). In spite of the broad range of $u_{{\ast}{\mathrm{t}}}$, all dune forms in fig. \ref{fig:Mars_bimodal_dunes} have been obtained with one single value of $u_{\ast} = 3.0$ m$/$s.} \label{tab:parameters_bimodal_dunes}
\end{table}

All dune shapes in fig. \ref{fig:Mars_bimodal_dunes} have been achieved with a time $T_{\mathrm{w}}$ in the range of $0.7$ to $5.8$ days. If the period is too large, of the order of a few months, then the dunes evolve into barchanoidal forms. 

The dune shape in fig. \ref{fig:Mars_bimodal_dunes}a$^{\prime}$ has been obtained with ${\theta}_{\mathrm{w}} = 100^{\circ}$ and with $T_{\mathrm{w}} = 250000$ s $\approx 2.9$ days, while ${\theta}_{\mathrm{w}} = 140^{\circ}$ and $T_{\mathrm{w}} = 500000$ s $\approx 5.8$ days has been used to calculate the dune in b$^{\prime}$. Moreover, we found that the structure observed in the dune field of fig. \ref{fig:Mars_bimodal_dunes}c can be obtained by a {\em{change}} in the local wind regime. The dune shape in fig. \ref{fig:Mars_bimodal_dunes}c$^{\prime}$ has been obtained in the following manner: (i) first, an elongated dune form as the one in fig. \ref{fig:Mars_bimodal_dunes}b$^{\prime}$ is formed with an angle ${\theta}_{\mathrm{w}} = 120^{\circ}$ and with $T_{\mathrm{w}} = 60000$ s $\approx 0.7$ day; (ii) next, the angle ${\theta}_{\mathrm{w}}$ has been reduced to $80^{\circ}$. Thereafter, the linear dune becomes unstable and decays into a string of rounded barchans as seen in fig. \ref{fig:Mars_bimodal_dunes}c. 

It is interesting to notice that our calculations provide a different explanation for the formation of the Martian dune field in fig. \ref{fig:MOC_images}e than that proposed by Bourke \cite{Bourke_2006}. We found that the field in fig. \ref{fig:MOC_images}e consists of linear dunes which are decaying into barchans, while Bourke \cite{Bourke_2006} suggested an alternative view: the small barchans would merge to form the linear dunes.

The results of fig. \ref{fig:Mars_bimodal_dunes} provide evidence for bimodal wind regimes on Mars. We find that a variety of Martian dune forms which appear in craters and which develop on bedrock have been formed by a wind whose direction alternates between two main orientations. We conclude that if more sand were available in those places, longitudinal dunes would in fact appear as observed in terrestrial sand seas. The study of the shape of linear dunes is the subject of a future publication \cite{Parteli_et_al_preprint}.

Again, the wind strength $u_{\ast} = 3.0$ m$/$s used in the calculations must be interpreted as the representative value of shear velocity that is above the threshold for saltation and is responsible for the major changes of the surface \cite{Sullivan_et_al_2005}. Because Martian winds are most of the time {\em{below}} the threshold for saltation, we expect the timescale $T_{\mathrm{real}}$ of the changes in wind directions to be in reality much larger than the values of $T_{\mathrm{w}}$ (a few days) found in the calculations. 

We define as $f_{\mathrm{w}}$ the fraction of time the wind strength is above the threshold for saltation. From the results of the calculations of barchans, this means that the Martian $u_{\ast}$ is around $3.0$ m$/$s a fraction $f_{\mathrm{w}}$ of time. We interpret this value of shear velocity as the representative wind friction speed associated with the gusts of aeolian activity that occur during the strongest dust storms on Mars \cite{Moore_1985}. Further, the real timescale $T_{\mathrm{real}}$ of the changes in wind direction is defined through the relation $f_{\mathrm{w}} = T_{\mathrm{w}}/T_{\mathrm{real}}$.

Let us assume that winds above the threshold on Mars occur generally during ${\Delta}t_{\mathrm{saltation}} = 40$ s at intervals of ${\Delta}T = 5$ years (2000 days or $1.728 \cdot 10^8$ s) \cite{Arvidson_et_al_1983,Moore_1985}, i.e. $f_{\mathrm{w}} = {\Delta}t_{\mathrm{saltation}}/{\Delta}T \approx 2.31 \cdot 10^{-7}$. A characteristic time $T_{\mathrm{w}} \approx 1 - 5$ days means $T_{\mathrm{w}} = 86,400 - 432,000$ seconds. Dividing $T_{\mathrm{w}}$ by ${\Delta}t_{\mathrm{saltation}} = 40$ s, this characteristic time corresponds to $2,160 - 10,800$ gusts of saltation transport. The Martian real time $T_{\mathrm{real}}$ is 
\begin{equation}
T_{\mathrm{real}} = \frac{T_{\mathrm{w}}}{{\Delta}t_{\mathrm{saltation}}} \times 5.0 \ \ {\mbox{years}} \approx 10,800 - 54,000 \ \ {\mbox{years}}. \label{eq:Treal_Mars}
\end{equation}
Therefore, the real time of oscillation of the wind direction on Mars found from our calculations is of the order of $10^4$ years, where it has been assumed that Martian saltation occurs as frequently as observed from the Mars Missions \cite{Arvidson_et_al_1983,Moore_1985,Sullivan_et_al_2005}.

\section{\label{sec:conclusions}Conclusions}

In the present work, we have applied a well established dune model, which successfully reproduces the shape of terrestrial dunes measured in the field, to study dune formation on Mars. In summary, we found that dunes observed in the images of Mars could have been formed by the action of sand-moving winds that occur occasionally under the present atmospheric conditions of the red planet. Below we give a list of the main conclusions:
\begin{itemize}
\item the quantities controlling Martian saltation, as the average grain velocity, mean saltation height and saturated flux may vary in a significant manner depending on the location on Mars. This is because local average values of Martian surface pressure and temperature may be very different depending on the geographical location;
\item from the shape of barchan dunes on Mars, we found that the rate at which Martian grains enter saltation is 10 times higher than on Earth. The Martian higher entrainment rate, which is a result of the larger splash events on Mars, shortens the length of flux saturation and reduces the scale of dunes that is obtained if only the Martian larger saltation length is considered \cite{Hersen_et_al_2002,Kroy_et_al_2005};  
\item all dune shapes studied in this work could be reproduced with values of shear velocity that do not exceed $u_{\ast} = 3.0 \pm 0.1$ m$/$s, independently of the location on Mars. We interpret this value as the representative friction speed of sand-moving winds that occur during the strongest dust storms on Mars;
\item for the same value of relative wind velocity $u_{\ast}/u_{{\ast}{\mathrm{t}}}$, barchans would move ten times faster on Mars than on Earth. However, the migration velocity of Martian barchans is negligible because saltation transport in fact occurs only seldom on the present Mars;
\item we found Martian dune shapes that have been formed by bimodal wind regimes. The timescale of changes in wind direction obtained in calculations is of the order of a few days. Taking into account that winds transport sand on Mars during some tens of seconds in intervals of a few years \cite{Arvidson_et_al_1983}, this timescale is in reality of the order of $10,000-50,000$ terrestrial years.
\end{itemize}

It is interesting to notice that a significant change in wind direction (by $90^{\circ}$ or more) is expected to occur after each extreme of the {\em{orbital}} cycle of Mars, which is determined by the combined effect of the precession of its axis and the arrival at {\em{perihelion}} \cite{Arvidson_et_al_1979,Thomas_1982,Lee_and_Thomas_1995,Fernandez_1998,Malin_et_al_1998,Thomas_et_al_1999}. Because of precession of the Martian axis, each pole of Mars appears tilted to the sun in a cycle of $51,000$ years. Now the latitude which ``looks'' toward the sun at perihelion is $15^{\circ}$S, but this ``subsolar latitude at perihelion'' (SLP) migrates $\pm 25^{\circ}$ about the Equator over a 51kYr time span \cite{Arvidson_et_al_1979}. This orbital cycle is the most important one for the climate of Mars, the hemisphere of the SLP having short, hot summers, and being the one of the major dust storms activity. In $25,500$ years, it is the northern hemisphere that will be tilted to the sun. ``{\em{The large amounts of fine dust currently deposited in the northern hemisphere in regions such as Tharsis, Arabia, and Elysium will be redistributed to the southern hemisphere}}'' \cite{Sheehan_1996}. This half cycle of $25,500$ years is in fact well within the range of characteristic time $10,800 < T_{\mathrm{real}} < 54,000$ years of bimodal winds found from our calculations of sand dunes on Mars.

In comparison, the timescale $T_{\mathrm{real}}$ of changes in directions of bimodal winds in terrestrial dune fields is of the order of a few weeks or months. On Earth, linear dunes appear due to wind changes that occur {\em{seasonally}} \cite{Livingstone_1989,Tsoar_1983}, since the fraction of time $f_{\mathrm{w}}$ the wind friction speed $u_{\ast}$ is above the threshold $u_{{\ast}{\mathrm{t}}}$ is much larger than on Mars \cite{Fryberger_and_Dean_1979,Tsoar_2005}. In this manner, we do not expect the wind changes due to precession of the Earth's axis to play a major role for the shape of terrestrial bimodal sand dunes. On the other hand, Martian winds that are not associated with the intense dust storms mentioned in the last paragraph are too weak to move sand. Indeed, such weak winds are responsible for the appearance of dust devils that leave ephemeral marks on the surface of Martian dunes \cite{Fenton_et_al_2003}, which appear unmobile.

The shape of Martian dunes could be only achieved with real wind and atmospheric conditions of the present Mars because the entrainment rate of grains into saltation, which we found to be 10 times higher on Mars than on Earth, was incorporated in the model equations. In fact, it is well known from experiments that the splash events on Mars are much larger than on Earth due to the higher velocity of Martian grains \cite{Marshall_et_al_1998}. What we have found in the calculations is the implication of the larger amount of splashed grains on Mars for the flux saturation and formation of dunes. It would be interesting to make a full microscopic simulation for the saltation mechanism of Mars similar to the one that was recently performed by Almeida {\em{et al.}} \cite{Almeida_et_al_2006} to confirm our findings microscopically.

\acknowledgments
We acknowledge Orencio Dur\'an for the numerous discussions and for his contribution at the initial stage of this work. Volker Schatz, Haim Tsoar and Kenneth Edgett are also acknowledged for discussions. We thank G\"unter Wunner, Harald Giessen, Jason Gallas and Adriano Sousa for the suggestions and many valuable comments. This research was supported in part by a Volkswagenstiftung and The Max-Planck Prize. E. J. R. Parteli acknowledges support from CAPES - Bras\'{\i}lia/Brazil.

\end{document}